\newif\ifconfver
\confvertrue        

\newif\ifonecoltab
\onecoltabtrue        

\newif\ifplainver  

\ifplainver
    \confverfalse                         
\fi

\ifconfver
     \documentclass[10pt,journal]{IEEEtran}
\else
    \ifplainver
        \documentclass[11pt]{article}
        \usepackage{fullpage}
    \else
        \documentclass[12pt,draftcls,onecolumn]{IEEEtran}
    \fi
\fi

\usepackage{calc,amsfonts,amssymb,amsmath,bm,url,color,theorem,graphicx,cite,epstopdf,nicefrac,stmaryrd}
\usepackage{psfrag,subfigure,float,epstopdf}
\usepackage[algoruled,linesnumbered]{algorithm2e}
\usepackage{multirow}
\usepackage{algorithmic}
\usepackage[normalem]{ulem}
\usepackage{mdframed}

\definecolor{orange}{RGB}{255,107,0}

\newtheorem{Lemma}{Lemma}
\newtheorem{Prop}{Proposition}
\newtheorem{Theorem}{Theorem}

\theoremstyle{definition}

\newtheorem{Remark}{Remark}

\newcommand{\W}{\boldsymbol{W}}
\newcommand{\F}{\boldsymbol{F}}
\newcommand{\D}{\boldsymbol{D}}
\newcommand{\I}{\boldsymbol{I}}
\newcommand{\Y}{\boldsymbol{Y}}
\newcommand{\G}{\boldsymbol{G}}
\newcommand{\Q}{\boldsymbol{Q}}
\newcommand{\X}{\boldsymbol{X}}
\newcommand{\C}{\boldsymbol{C}}

\newcommand{\U}{\boldsymbol{U}}
\newcommand{\V}{\boldsymbol{V}}

\newcommand{\one}{\boldsymbol{1}}
\newcommand{\zero}{\boldsymbol{0}}
\renewcommand{\H}{\boldsymbol{H}}
\newcommand{\M}{\boldsymbol{M}}
\newcommand{\A}{\boldsymbol{A}}
\newcommand{\MP}{\boldsymbol{P}}
\newcommand{\B}{\boldsymbol{B}}
\renewcommand{\S}{\boldsymbol{S}}

\newcommand{\x}{\boldsymbol{x}}
\newcommand{\z}{\boldsymbol{z}}

\newcommand{\y}{\boldsymbol{y}}

\newcommand{\T}{{\!\top\!}}

\newcommand{\bPi}{\bm{\Pi}}
\newcommand{\bpi}{\bm{\pi}}

\newcommand{\Diag}{\text{Diag}}
\newcommand{\tvec}{\text{vec}}

\newcommand{\tX}{\underline{\bm X}}
\newcommand{\tN}{\underline{\bm N}}

\newcommand{\tW}{\underline{\bm W}}
\newcommand{\tY}{\underline{\bm Y}}

\def \bLambda {\boldsymbol{\Lambda}}
\def \tvec {\text{vec}}

\def \tDiag {\text{Diag}}
\def \dd {^{(d)}}
\def \dz {^{(d_0)}}

\def \id {^{(i_d)}}
\def \idp {^{(i_{d+1})}}

\def \vc {\boldsymbol{c}}
\def \CI {\mathcal{I}}
\def \CJ {\mathcal{J}}
\def \CK {\mathcal{K}}
\def \CP {\mathcal{P}}
\def \CQ {\mathcal{Q}}
\def \CR {\mathcal{R}}
\def \vd {\boldsymbol{d}}
\def \io {^{(i_1)}}

\newcommand{\bbR}{\mathbb{R}}

\def \cS { {\mathcal{S}} }
\def \R {\bm{R}}

\DeclareMathOperator*{\minimize}{\textrm{minimize}}

\def \io {^{(i_1)}}

\begin{document}

\newcommand{\papertitle}{
Spectrum Cartography via Coupled Block-Term Tensor Decomposition
}

\newcommand{\paperabstract}{%
Spectrum cartography aims at estimating power propagation patterns over a geographical region across multiple frequency bands (i.e., a \textit{radio map})---from limited samples taken sparsely over the region. Classic cartography methods are mostly concerned with recovering the aggregate radio frequency (RF) information while ignoring the constituents of the radio map---but fine-grained emitter-level RF information is of great interest. In addition, many existing cartography  methods  explicitly or implicitly assume random spatial sampling schemes that may be difficult to implement, due to legal/privacy/security issues. The theoretical aspects (e.g., identifiability of the radio map) of many existing methods are also unclear.
In this work, we propose a joint radio map recovery and disaggregation method that is based on coupled block-term tensor decomposition. Our method guarantees identifiability of the \textit{individual} radio map of \textit{each emitter} (thereby that of the aggregate radio map as well), under realistic conditions. The identifiability result holds under a large variety of geographical sampling patterns, including a number of pragmatic systematic sampling strategies. We also propose effective optimization algorithms to carry out the formulated radio map disaggregation problems. Extensive simulations are employed to showcase the effectiveness of the proposed approach.
}


\ifplainver

    \date{\today}

    \title{\papertitle}

    \author{
    Xiao Fu$^\ast$
    \\ ~ \\
		$^\ast$School of Electrical Engineering and Computer Science\\ Oregon State University\\
$^\dag$Department of Systems Engineering and Engineering Management\\ The Chinese University of Hong Kong\\
$^\star$Department of Computer and Information Science and Engineering\\ University of Florida\\
	  \\~
    }

    \maketitle

    \begin{abstract}
    \paperabstract
    \end{abstract}

\else
    \title{\papertitle}

    \ifconfver \else {\linespread{1.1} \rm \fi

\author{Guoyong Zhang, Xiao Fu, \IEEEmembership{Member, IEEE}, Jun Wang, \IEEEmembership{Member, IEEE}\\ Xi-Le Zhao, \IEEEmembership{Member, IEEE}, Mingyi Hong, \IEEEmembership{Member, IEEE}
	
\thanks{

This work of X. Fu and M. Hong is supported in part by National Science Foundation under Project ECCS 1808159 and the Army Research Office under Project ARO W911NF-19-1-0247.
G. Zhang and J. Wang are supported in part by the National Key R\&D Program of China under Grant 2018YFC0807101, the National Research Program of China under Grant 9020302, the Foundation of National Key Laboratory of Science and Technology on Communications, the Innovation Fund of NCL (IFN), and the National Natural Science Foundation of China (NSFC) under Grant 61471099.

G. Zhang and J. Wang are  with National Key Laboratory of Science and Technology on Communications at University of Electronic Science and Technology of China (UESTC), Chengdu, China. emails: guoyong.zhang158@gmail.com and junwang@uestc.edu.cn.  The work is developed during G. Zhang's visit to  Oregon State University.
X. Fu is with the School of Electrical Engineering and Computer Science, Oregon State University (OSU), Corvallis, OR 97331, United States. email: xiao.fu@oregonstate.edu;
M. Hong is with the Department of Electrical and Computer Engineering, University of Minnesota, Minneapolis, MN 55455. email: mhong@umn.edu
X. Zhao is  with the School of Mathematical Sciences at University of Electronic Science and Technology of
China, Chengdu, China. e-mail: xlzhao122003@163.com

}
}

    \maketitle

    \ifconfver \else
        \begin{center} \vspace*{-2\baselineskip}
        \end{center}
    \fi

    \begin{abstract}
    \paperabstract
    \end{abstract}

    \begin{IEEEkeywords}\vspace{-0.0cm}%
        coupled tensor decomposition, tensor completion, block term decomposition, radio map, slab sampling, fiber sampling, spectrum cartography
    \end{IEEEkeywords}

    \ifconfver \else \IEEEpeerreviewmaketitle} \fi

 \fi

\ifconfver \else
    \ifplainver \else
        \newpage
\fi \fi

\section{Introduction}

Improving spectrum efficiency relies on accurate, fine-grained, and agile radio frequency (RF) awareness.
\textit{Spectrum sensing} is considered the first step towards RF awareness. Spectrum sensing is commonly { formulated} as a detection problem, which determines if a frequency band is used or not \cite{yucek2009survey}; some works also consider it as a power spectral density (PSD) estimation problem that recovers a wideband PSD using a sub-Nyquist sampling rate \cite{hamdaoui2018compressed,tian2007compressed}. \textit{Spectrum cartography} \cite{hamid2017non} has a more ambitious goal: It aims to construct the received signal power over some bands of interest and a geographical area---i.e., a `radio map' \cite{bi2019engineering}.

As radio maps can provide valuable spatial and spectral information, it can be utilized to enhance the performance of many classic wireless communications/networking tasks, e.g., opportunistic access, interference management, networking planning and spectrum surveillance; see a recent survey in \cite{bi2019engineering}. Radio maps also find applications in a number of emerging topics such as  indoor positioning \cite{zou2017winips} and optimal relay deployment for \textit{unmanned aerial vehicle} (UAV)-based networks \cite{chen2017learning,chen2017optimal}.

Many early approaches for radio map estimation focus on single-band radio map completion from sparsely sampled measurements over a geographical region.
This problem can be understood as an image inpainting problem, and spatial smoothness of the power { propagation} patterns is the main prior { information} exploited for handling the task. Based on this perspective, a number of approaches were proposed, e.g., the nonparametric methods (also called model-free methods) based on kernel regression, leveraging on different kernels such as Kriging  \cite{boccolini2012wireless}, thin plate splines (TPS) \cite{ureten2012comparison}, and Gaussian radial basis functions (RBF) \cite{hamid2017non}.
Some parametric methods based on the knowledge of power propagation models were also proposed, e.g., the sparse representation-based works as in \cite{jayawickrama2013improved,bazerque2009distributed,jayawickrama2014iteratively,huang2014cooperative}.
Estimating multi-band radio maps has also been considered in the literature \cite{bazerque2009distributed,bazerque2011group,romero2017learning}. The problem is much harder than the single-band case, since multi-band radio map is a third-order tensor---and tensor completion is a hard task.
In the literature, a multi-band radio map is often modeled as a superposition of \textit{emitter PSDs} scaled by their respective \textit{spatial loss fields} (SLFs).
Some prior knowledge of the SLFs and the PSDs is assumed to be known in \cite{bazerque2009distributed} and \cite{bazerque2011group,romero2017learning}, respectively, to simplify the tensor completion task.

Both spectrum sensing and cartography have made considerable { progress} since the early 2000s. However, RF awareness-enabling techniques are still far from satisfactory. First, most of the aforementioned techniques only recover aggregate PSD or SLF that are normally contributed by multiple transmitters. Simultaneously estimating PSD and SLF disaggregated ({ i.e., unmixed}) to the emitter level has not been addressed. Second, most existing cartography techniques put { emphasis} on the algorithmic aspects, e.g, interpolation and kernel design. However, it has been unclear if in theory the spectral and spatial information { are} recoverable, given limited samples over space and/or frequency. Third, most cartography methods exhibit the best performance under random spatial sampling (which is perhaps reminiscent of the effectiveness of matrix/tensor completion under randomly missed entries \cite{candes2009exact,liu2013tensor}), while random sampling may not always be realistic, because sensing in some areas may be subject to security/privacy/legal constraints.

\smallskip

\noindent
{\bf Contributions.}
In this work, we investigate the theoretical and practical aspects of radio map recovery and disaggregation under different measurement acquisition strategies.
We propose an analytical framework based on coupled tensor decomposition for radio map disaggregation from partial observations. Our framework is based on modeling the radio maps as tensors following the \textit{multilinear rank-$(L_r,L_r,1)$ block-term decomposition} (BTD) model (which will be referred { as} the {\sf LL1} decomposition model for simplicity) \cite{de2008decompositions2}.
Our formulations can be understood as joint {\sf LL1} decomposition of tensors with shared latent factors, wherein each tensor consists of measurements taken from the whole radio map.
We derive identifiability guarantees for the emitter PSDs and SLFs under this framework. Our identifiability results cover a variety of realistic measurement collection schemes, including systematic (non-random) sensor deployment strategies.
Note that guaranteed PSD and SLF recovery under systematic sensing schemes is desired in practice, since in many cases random sensor deployment may be hard.
On the { algorithmic} side, we propose a \textit{block coordinate descent} (BCD)  framework to handle the formulated joint PSD and SLF recovery problems.
Extensive simulations corroborate our analysis and show the usefulness of the proposed approach under realistic scenarios with challenging fading and shadowing effects.

We should mention that there { is} a number of works, such as \cite{romero2017learning} and \cite{fu2015factor,fu2016power}, which also estimate emitter-level information. However, the work in \cite{romero2017learning} assumes that the PSD of each emitter is known, which may not be the case in practice. { In addition, the approaches in \cite{fu2015factor,fu2016power} do not consider the SLF recovery problem.}
A conference version of this work was presented at Asilomar 2019 \cite{zhang2019coupled}. The journal version additionally includes more sampling patterns and their associated recovery guarantees, detailed { recoverability} proofs, algorithm design, and comprehensive numerical study.

\smallskip

\noindent
{\bf Notation.} We largely follow the established convention in signal processing. In particular, $x$, $\x$, $\X$, and $\tX$ denote a scalar, a vector, a matrix, and a tensor, respectively. $\tX(i,j,k)$ (or $x_{i,j,k}$, $(\tX)_{i,j,k}$) denotes the $(i,j,k)$th element of a third-order tensor. $c_{k,r}$ (or $\C(k,r)$) denotes the $(k,r)$th element of a matrix. $c_k$ (or $\vc(k)$) denotes $k$th element of a vector. The superscripts $^\T$ and $^\dag$ denote the transpose and the pseudoinverse operations of a matrix, respectively.  We use the MATLAB notations to represent a submatrix of a given matrix or subtensor of a given tensor; e.g., $\C(:,r)$ denotes $r$-th column of $\C$, and $\tX(:,:,k)$ denotes $k$-th frontal slab of $\tX$. { The function $\lfloor x\rfloor$ denotes the largest integer less than or equal to $x$}. The cardinality of set $\CK$ is denoted by $\vert\CK\vert$. $[K]$ denotes the set $\{1,\cdots,K\}$. The outer product `$\circ$' between a matrix $\S$ and a vector $\vc$ is defined as $\tX=\S\circ\vc$ with $(i,j,k)$th element $\tX(i,j,k)=\S(i,j)\vc(k)$. The $L_1$-norm of a vector and matrix are defined as $\Vert\x\Vert_1=\sum_i \vert x_i\vert$ and $\Vert\X\Vert_1=\sum_{ij} \vert x_{ij}\vert$, respectively. We use $\Vert\X\Vert_F=\sqrt{\sum_{ij}x_{ij}^2}$ and $\Vert\tX\Vert_F=\sqrt{\sum_{ijk}x_{ijk}^2}$ to denote the Frobenius norms of $\X$ and $\tX$, respectively. { The \textit{Khatri-Rao product} (column-wise Kronecker) of two matrices $\X\in\bbR^{I\times K}$ and $\Y\in\bbR^{J\times K}$ is defined as
$
\X\odot\Y=[\x_1\otimes\y_1,\cdots,\x_K\otimes\y_K],$
where $\otimes$ denotes the Kronecker product. }

\section{Problem Statement}
We consider a scenario where $R$ emitters exist within a 2-D geographical area of interest. Each emitter transmits { across}  certain frequency bands, and the frequency bands of different emitters could overlap.
To be more precise, assuming that the signals from different emitters (or, sources) are uncorrelated, the received aggregated \textit{power spectral density} (PSD) at spatial coordinates $(i,j)$ and frequency $f_k$ is represented as \cite{bazerque2011group,romero2017learning}:
\begin{equation}\label{eq:sigmod}
x_{i,j,k} \approx \sum_{r=1}^R \S_r(i,j)c_{k,r},
\end{equation}
where $\S_r\in\mathbb{R}^{I\times J}$ denotes the (discretized) SLF of source $r$ over the geographical region of interest,
$c_{k,r}$ denotes the PSD of source $r$ measured at frequency $f_k$,
and $R$ the number of sources.
The signal model in \eqref{eq:sigmod} means that the received power across frequencies at each geographical point in space is a superposition of the PSD of different sources---scaled by their respective SLFs. Eq.~\eqref{eq:sigmod} is a valid model if the bandwidth of interest is relatively small compared to the carrier frequency \cite{fu2015factor,fu2016power,romero2017learning} (e.g., if the band of interest spans 20MHz at a carrier frequency within 2-5 GHz---and the bandwidth can be even larger when the carrier frequency is at the range of millimeter wavelengths).

It is natural to represent the received space-space-frequency signal using tensor notations. Let us define a tensor $\tX\in\mathbb{R}^{I\times J\times K}$ and a matrix $\C\in\mathbb{R}^{K\times R}$ such that
$    \tX(i,j,k)=x_{i,j,k}$ and  $\C(k,r)=c_{k,r}$.
Then, if noise is absent, the signal model can be expressed as{{:}}
\begin{equation}\label{eq:sigmod_tensor}
\tX = \sum_{r=1}^R \S_r \circ \vc_r,
\end{equation}
where $\vc_r$ denotes the $r$th column of $\C$ and $\circ$ denotes the outer product.
We refer to the tensor $\tX$ as a {\it radio map}, since it reveals the RF environment across both the spatial and the spectral domains.
One remark is that the RF environment can change quickly in practice, and thus one may also add a temporal dimension to make the above signal model a fourth-order tensor.
Nonetheless, if $\tX$ is measured as geographical \textit{power spectral density}, it is expected to reflect the averaged prominent RF situation over a relatively long period---which is normally informative enough.
Hence, in this work, we focus on the spatial-spectral case.

%
%

To put into context, our problem is to estimate $\bm S_r$ for $r=1,\ldots,R$ and $\bm C$ given \[\tX(i,j,k),\quad (i,j,k)\in{\bm \Omega},\] and $\bm \Omega$ is the index set of observed entries of $\tX$. Note that normally $|{\bm \Omega}|\ll IJ{ K}$---i.e., the RF tensor is heavily undersampled. { In this work, we refer to the problem of estimating the individual $\bm c_r$ and $\S_r$ as RF tensor \textit{disaggregation} from limited observations---as opposed to approaches that only recover the manifest/aggregated information $\tX$. Obviously, if the former can be properly done, the latter is trivial to recover.}

Note that if one only aims at recovering $\tX$, the problem is essentially a {\it tensor completion problem} under a certain observation model. Many approaches for this problem appeared in the literature \cite{liu2013tensor,cong2011sparse,xu2013block,hu2013fast,goldfarb2014robust,zhang2017exact,yuan2016tensor,mu2014square}, mainly for image inpainting, denoising, and other vision problems. Nevertheless, many of these tensor completion approaches (especially those who admit theoretical supports \cite{yuan2016tensor,mu2014square,zhang2017exact}) work under an assumption that the missing/observed values happen uniformly at random over the tensor. Translating this to our problem, it would require that many sensors are randomly deployed all over the geographical region of interest and that the sensors are randomly sensing different frequency bands. In some cases, it is hard to fulfill these requirements, especially when sensing is conducted by some secondary (or unlicensed) systems.
Sensor deployment has many restrictions in practice (e.g., due to privacy/legal/security issues), and can hardly be random.  
This suggests systematic sampling strategies may be more appropriate. 
More importantly, tensor completion can only recover the ambient data $\tX$, but not the constituents $\bm S_r$ and $\bm C$---which represent the fine-grained information.
In this work, we will offer solutions for radio map disaggregation, in particular, under systematic sampling schemes.

%
%

\section{Preliminaries: Block Decomposition Into Multilinear Rank-$(L_r,L_r,1)$ Terms}
To better understand our approach, we first introduce some relevant notions and terminologies in tensor analytics.
A tensor $\tX$ is a multidimensional array.
 Unlike matrices whose rank is mathematically well-defined, the definition of tensor rank is nonsingular \cite{sidiropoulos2017tensor,kolda2009tensor}.
Some popular tensor decomposition models such as the \textit{canonical polyadic decomposition} (CPD) (previously known as \textit{parallel factor analysis} (PARAFAC) \cite{harshman1994parafac}) and the Tucker decomposition \cite{tucker1966some} have already triggered a large number of applications in sensing and communications \cite{sidiropoulos2000parallel,sidiropoulos2001identifiability,fu2016power,rong2005blind,vorobyov2005robust}.

\subsection{Decomposition in Multilinear Rank-$(L_r,L_r,1)$ Terms}
\begin{figure}[t]
	\centering
	\includegraphics[width=0.8\linewidth]{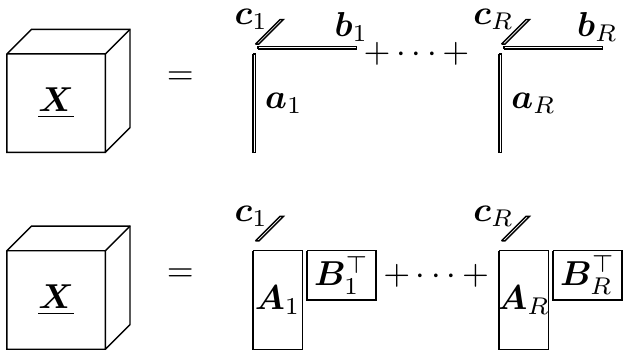}
	\caption{Illustration of tensor decomposition models. Top: CPD. Bottom: $(L,L,1)$-BTD.}
	\label{fig:cpdbtd}
\end{figure}

Beyond the CPD and Tucker models, yet another very useful tensor model is the so-called \textit{block term decomposition} (BTD) model \cite{de2008decompositions1,de2008decompositions2,de2008decompositions3}.
BTD subsumes many decomposition models as its special cases and thus is quite general. In particular, we are interested in a special kind of BTD, which is sometimes referred to as \textit{tensor decomposition in multilinear rank-$(L_r,L_r,1)$ terms}, or, simply the {\sf LL1} decomposition model \cite{de2011blind}. To be specific, under the {\sf LL1} model, a third-order tensor can be written as \cite{de2011blind,de2008decompositions2}:
\begin{equation}\label{eq:LL1}
\tX = \sum_{r=1}^R (\A_r\B_r^\top)\circ \vc_r,
\end{equation}
where $\A_r\in\mathbb{R}^{I\times L_r}$ and $\B_r\in\mathbb{R}^{J\times L_r}$ for $r=1,\ldots,R$, $\C=[\vc_1,\cdots,\vc_R]\in\mathbb{R}^{K\times R}$, and `$\circ$' denotes the outer product as before.
In this model, the tensor is a sum of outer products of a rank-$L_r$ matrix (we have ${\rm rank}(\A_r\B_r^\T)=L_r$ if $L_r\leq \min\{I,J\}$ and $\A_r$ and $\B_r$ have full column rank) and a vector.
The block term $ (\A_r\B_r^\top)\circ \vc_r$ is a tensor, which is said to have a {\it multilinear rank} of $(L_r,L_r,1)$ \cite{de2008decompositions1,de2008decompositions2,de2008decompositions3}. CPD is a special case of this model whose block terms have a multilinear rank of $(1,1,1)$;
see Fig.~\ref{fig:cpdbtd} for the difference between the popular CPD model and the {\sf LL1} model.
Also see the relationship between the BTD and Tucker decomposition models in \cite{de2008decompositions1}.

\subsection{Mode-$n$ Unfolding}
\begin{figure}[t]
	\centering
	\includegraphics[width=0.8\linewidth]{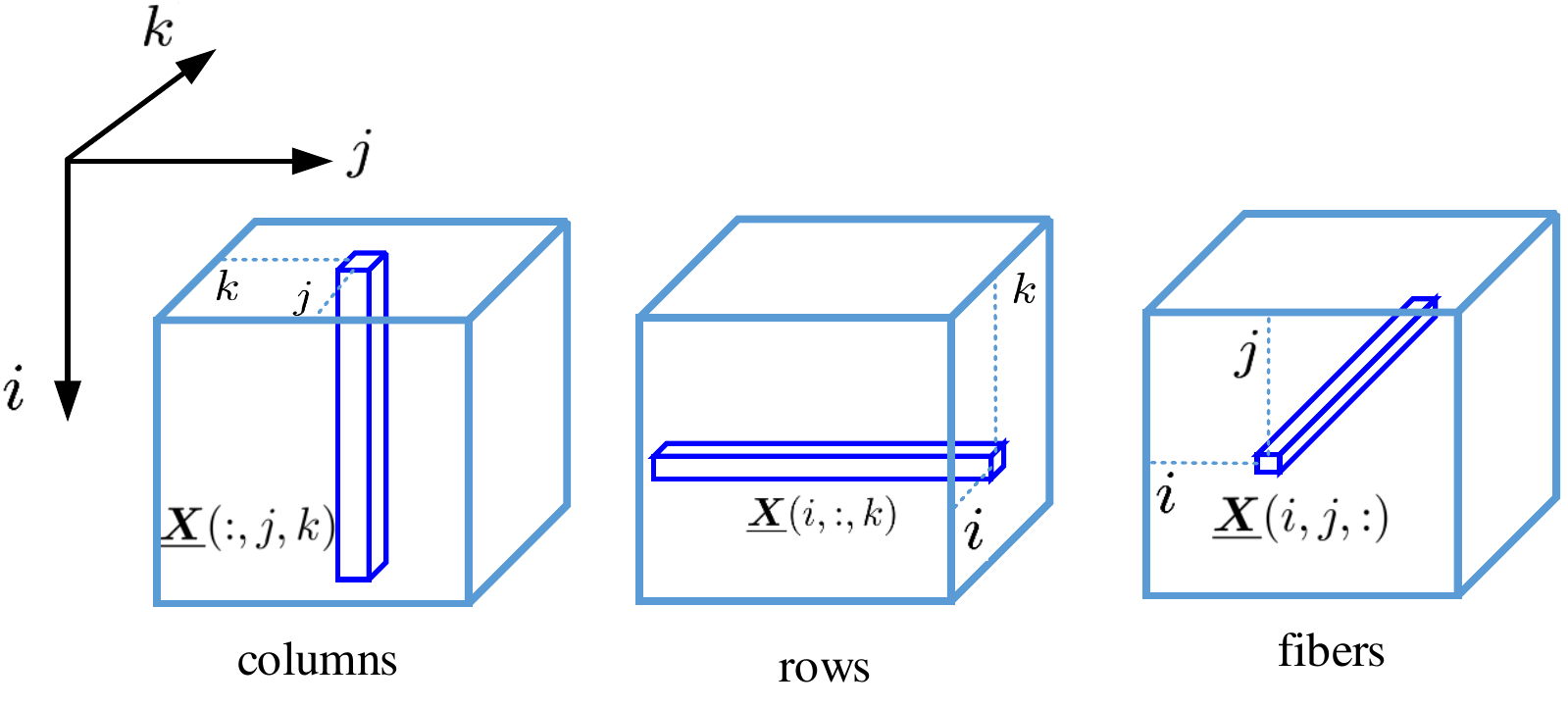}
	\caption{Illustration of three modes, i.e., column $\tX(:,j,k)$, row $\tX(i,:,k)$ and fiber $\tX(i,j,:)$, of a third-order tensor.}
	\label{fig:modes}
\end{figure}

\begin{figure}[t]
	\centering
	\includegraphics[width=0.8\linewidth]{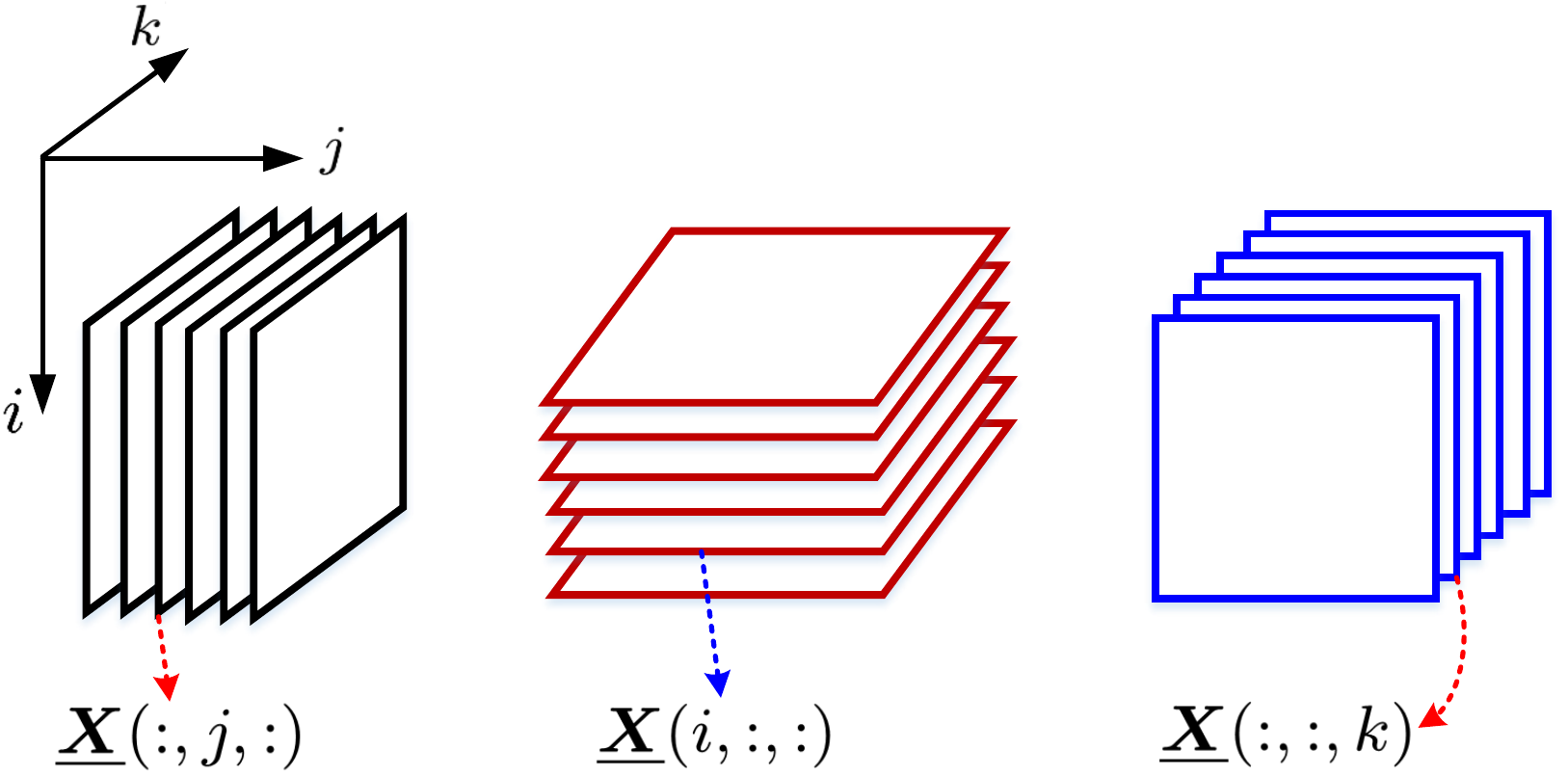}
	\caption{Illustration of the vertical slabs $\tX(:,j,:)$ (left), the horizontal slabs $\tX(i,:,:)$ (middle), and the frontal slabs $\tX(:,:,k)$ (right) of a third-order tensor.}
	\label{fig:slabs}
\end{figure}

A third-order tensor $\tX\in\bbR^{I\times J\times K}$ admits three modes:  rows $\tX(i,:,k)$, columns $\tX(:,j,k)$, fibers $\tX(i,j,:)$---see Fig.~\ref{fig:modes}. There are also three types of slabs associated with a third-order tensor, namely, the horizontal slabs $\tX(i,:,:)$, the vertical slabs $\tX(:,j,:)$ and the frontal slabs $\tX(:,:,k)$, respectively---see Fig.~\ref{fig:slabs}.

A tensor can be represented in a matrix form by unfolding { (or matricizing) the tensor}. For example the mode-$1$ unfolding of $\tX$, denoted as $\X_1$, can be obtained by stacking all the vectorized horizontal slabs $\tX(i,:,:)$ as follows:
\begin{align*}
\X_1 &= [\tvec(\tX(1,:,:)),\cdots,\tvec(\tX(I,:,:))]\\
&=(\C\odot_p\B)\A^\top,
\end{align*}
where $\tvec(\cdot)$ is the vectorization operator, and { the notation $\odot_p$ stands for the so-called partition-wise Khatri-Rao product, i.e., $
\C\odot_p\A = [\vc_1\otimes\A_1,\cdots,\vc_R\otimes\A_R]\in\bbR^{IK\times \sum_{r=1}^RL_r}$, in which $\A=[\A_1,\cdots,\A_R]\in\bbR^{I\times \sum_{r=1}^RL_r}$}. The mode-$2$ and mode-$3$ unfoldings (denoted as $\X_2$ and $\X_3$) are defined in a similar way as follows:
\begin{subequations}\label{Xunfold}
\begin{align}
\X_2& = [\tvec(\tX(:,1,:)),\cdots,\tvec(\tX(:,J,:))]\nonumber\\
&=(\C\odot_p\A)\B^\top,\\
\X_3& = [\tvec(\tX(:,:,1)),\cdots,\tvec(\tX(:,:,K))]\nonumber\\
&=[(\B_1\odot\A_1)\one_{L_1},\cdots,(\B_R\odot\A_R)\one_{L_R}]\C^\top, \nonumber\\
	& = [\tvec(\A_1\B_1^\top),\cdots,\tvec(\A_R\B_R^\top)]\C^\top.
\end{align}
\end{subequations}
In particular, note that one can re-write
\begin{equation}
\X_3 =\S\C^\T,
\end{equation}
where $\S(:,r)=\tvec(\A_r\B_r^\top)$. For more details, see \cite{de2008decompositions1,de2008decompositions2,de2008decompositions3,sidiropoulos2017tensor} ({ in particular, our unfolding strategy follows the paradigm in \cite{sidiropoulos2017tensor}}).
These representations will be handy in our theoretical analysis and algorithm design.

\subsection{Mode Product and BTD Representation}
Consider $\tX\in\bbR^{I\times J\times K}$ and $\MP_1\in\bbR^{M\times I},\MP_2\in\bbR^{N\times J},\MP_3\in\bbR^{T\times K}$.
The mode-$n$ product $\underline{\G}_n=\tX\times_n \MP_n$ for $n=1,2,3$ such that $\underline{\G}_1({m,j,k})=\sum_{i=1}^I\tX(i,j,k)\MP_1(m,i)$,
$\underline{\G}_2({i,n,k})=\sum_{j=1}^J\tX(i,j,k)\MP_2(n,j)$, and $\underline{\G}_3({i,j,t})=\sum_{k=1}^K\tX(i,j,k)\MP_3(t,k)$.
A key observation is that, if
$\tX$ follows the {\sf LL1} decomposition model as in \eqref{eq:LL1}, we have
\[
\tX\times_1\MP_1\times_2\MP_2\times_3\MP_3=
 \sum_{r=1}^R (\MP_1\A_r(\MP_2\B_r)^\top)\circ (\MP_3\vc_r).
\]

\subsection{Essential Uniqueness}
The {\sf LL1} model has a very nice property---$\{\A_r\B_r^\top\}_{r=1}^R$ and $\{\vc_r\}_{r=1}^R$  are identifiable up to permutation and scaling ambiguities, under some mild conditions.
In the model (\ref{eq:LL1}), permutation and scaling ambiguities mean that one can arbitrarily permute the terms $\{(\A_r\B_r^\T)\circ\vc_r\}_{r=1}^R$, and scale and counter-scale $\A_r\B_r^\T$ and $\vc_r$ as long as their outer product remains uncharged. Note that $\A_r\B_r^\top=\A_r\F_r(\B_r\F_r^{-\top})^\top$ holds for any non-singular matrix $\F_r\in\bbR^{L_r\times L_r}$. Hence, $\A_r$ and $\B_r$ cannot be identified from the {\sf LL1} decomposition of $\tX$ in general.
In this paper, we will use the following theorem:
\begin{Theorem} \cite{de2008decompositions2} \label{BTD_theorem_IJKR}
	Let $\tX=\sum_{r=1}^R(\A_r\B_r^\T )\circ \bm c_r$ and assume $L_r=L$ for $r=1,\cdots,R$. Assume that $(\A,\B,\C)$ are drawn from any joint absolutely continuous distributions. Suppose the following condition holds:
\begin{equation*}
K\ge R \;\text{and}\; \min\left(\bigg\lfloor\frac{I}{L}\bigg\rfloor,R\right) + \min\left(\bigg\lfloor\frac{J}{L}\bigg\rfloor,R\right)\ge R+2.
\end{equation*}
Denote $\S=[{\rm vec}(\S_1),\ldots,{\rm vec}(\S_R)]\in\mathbb{R}^{ IJ\times R}$ where $\S_r=\A_r\B_r^\T$. Then, with probability one, for any $\widetilde{\S}\in\mathbb{R}^{ IJ\times R}$ and $\widetilde{\C}\in\mathbb{R}^{K\times R}$ satisfying $\X_3 = \widetilde{\S}\widetilde{\C}^\T$, we have
\[       \widetilde{\C} =\C\bm \Pi\bm \Lambda,~     \widetilde{\S}   =\S\bPi\bLambda^{-1},   \]
where $\bm \Pi$ is a permutation matrix and $\bm \Lambda$ is a nonsingular diagonal matrix; i.e., the multilinear rank-$(L_r,L_r,1)$ decomposition of $\tX$ is essentially unique almost surely.
\end{Theorem}

Many more conditions for the essential uniqueness of BTD were proposed in \cite{de2008decompositions2,de2011blind}---e.g., the following one that considers the case where $K\geq R$ is relaxed:
\begin{Theorem} \cite{de2008decompositions2} \label{BTD_theorem_noC}
Let $\tX=\sum_{r=1}^R(\A_r\B_r^\T )\circ \bm c_r$ and assume $L_r=L$ for $r=1,\cdots,R$.
Assume $(\A,\B,\C)$ are drawn from certain joint absolutely continuous distributions. If $IJ\ge L^2R$ and
\[
\min\left(\bigg\lfloor\frac{I}{L}\bigg\rfloor,R\right) + \min\left(\bigg\lfloor\frac{J}{L}\bigg\rfloor,R\right)+\min(K,R)\ge 2R+2,
\]
then, the multilinear rank-$(L_r,L_r,1)$ decomposition of $\tX$ is essentially unique almost surely.
\end{Theorem}

{ For notational simplicity, we will by default use $L_1=\ldots =L_R=L$ in the sequel.}

\section{Proposed Approach}
In this section, we propose a framework for radio map estimation and disaggregation using down-sampled data.
Our method provably works with both {\it systematic} and \textit{random} sensor deployment strategies. For the former one, the sensing { patterns} are under control of the system designers.
Another salient feature of our approach is that it guarantees identifiability of the SLFs and power spectra for all the emitters in the region of interest---i.e., it allows emitter-level radio map reconstruction---under some conditions.

\subsection{Signal Model}
Our idea is to connect the radio map signal model in \eqref{eq:sigmod_tensor} with the {\sf LL1} model in \eqref{eq:LL1}, and then use a judiciously designed decomposition criterion to recover the latent factors of the heavily down-sampled tensor---thereby achieving the goal of disaggregation.

\begin{figure}
		\centering
	\includegraphics[width=.98\linewidth]{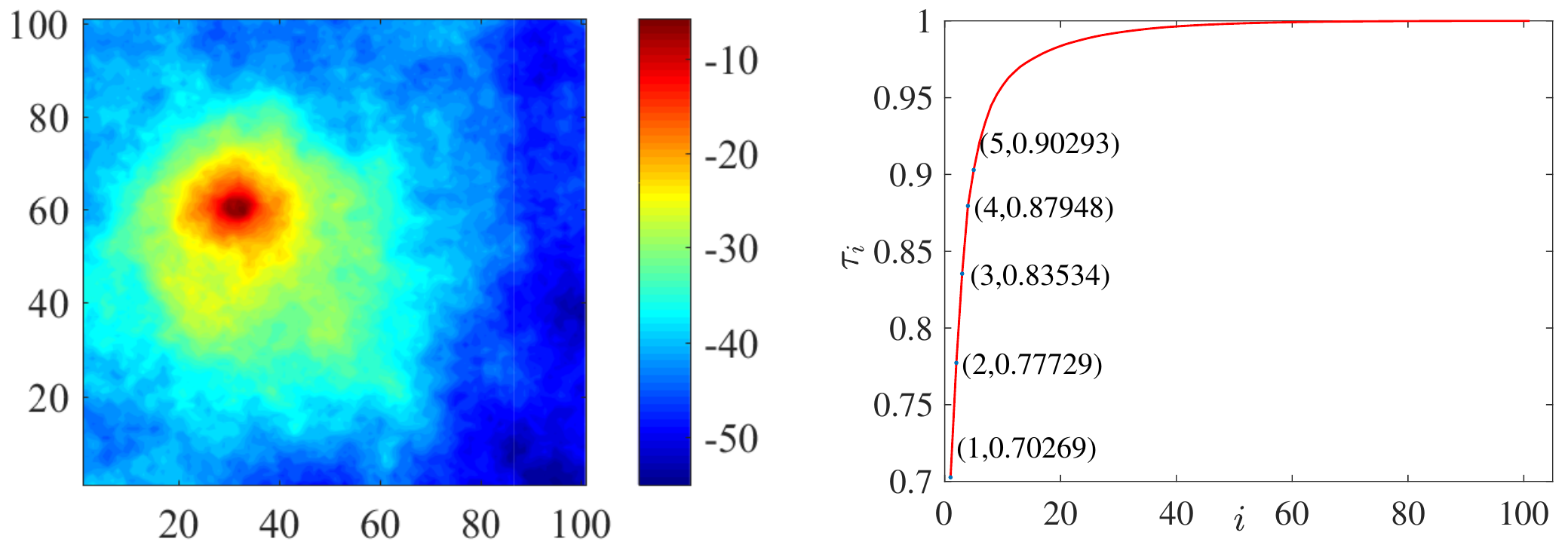}
	\caption{ Left: SLF of a source over a 100 m$^2$ region (discretized as a $101\times 101$ matrix), with shadowing effect. Right: The plot of $\tau_i$ versus $i$---showing that the SLF is approximately low-rank. The first 5 $i$-th index and its proportion $\tau_i$ is marked in the figure, namely $(i,\tau_i)$ for $i=1,2,3,4,5$.}
	\label{fig:low-rank}
\end{figure}

Our approach is motivated by the following {\it key postulate}: the SLF of a source $r$, denoted by $\S_{r}\in \mathbb{R}^{I\times J}$, is { approximately} low-rank for each individual source.  This assumption is reasonable, because of the continuity and smoothness of power propagation over space \cite{lee2017channel,alevizos2018limited,qian2018tensor,dupleich2015double}.
A simulated example can be found in Fig.~\ref{fig:low-rank}, where the SLF of a source over a 100 m $\times$ 100 m region is simulated under realistic shadowing effect following the \textit{log-normal shadowing model} with a standard deviation of 4 (which means a relatively strong shadowing effect) \cite[Chapter 4]{goldsmith2005wireless}.
We apply the singular value decomposition (SVD) to the SLF and plot the ratio
$
  \tau_i = \sum_{k=1}^i{ \mu_k}/\sum_{k=1}^{100} { \mu_k},
$
where ${ \mu_k}$ denotes the $k$th singular value of $\S_r$ in the figure.
From Fig.~\ref{fig:low-rank}, one can see $\tau_5=0.90293,$ which means that the top-5 rank-one components contain 90\% energy of $\S_r$.
This observation suggests that the SLFs, even under shadowing effects, may be well-approximated by low-rank matrices. { Using low-rank matrices to approximate spatial smoothness appeared in the literature, e.g., for channel gain field estimation \cite{lee2017channel} and fMRI processing \cite{chatzichristos2019blind}, but has not bee considered for SLFs, to our best knowledge}.

Under the low-rank postulate,
one  can express $\S_r$ by:
 $$\S_r=\A_r\B_r^\top,$$
where $\A_r\in\mathbb{R}^{ I\times L}$, $\B_r\in\mathbb{R}^{ J\times L}$ and ${ L}\ll\min\{I,J\}$. Consequently, we have
$$ \tX = \sum_{r=1}^R \S_r \circ \vc_r = \sum_{r=1}^R \left(\A_r\B_r^\top \right) \circ \vc_r; $$
i.e., the complete RF tensor follows a multilinear rank-${ (L,L,1)}$ BTD model.
Based on this model, we will propose a number of sensing and $(\S_r,\vc_r)$-estimation schemes in the following subsections.




\subsection{The Moving Sensor Case: Sensing Consecutive Grids}
If $\tX$ is available, then estimating $\S_r$ and $\vc_r$ can be done by directly applying {\sf LL1} algorithms (e.g., that in \cite{de2008decompositions3}) to the tensor.
In practice, however, the complete $\tX$ is not available. What we observe is a substantially undersampled version of $\tX$.
One possible route is to first apply tensor completion techniques to estimate the complete $\tX$, and then seek the ${\sf LL1}$ decomposition of the estimated $\tX$.
However, random sampling schemes that may ensure provable completion of $\tX$ may not be possible in spectrum cartography---since sensor may not be allowed acquire samples in many regions because of legal/security/privacy issues.

To handle the above challenges, consider that one employs a moving sampling device (e.g., a UAV or a regular/self-driving car) to sample an area of interest.  We assume that the spatial area is finely discretized into  $I\times J$ grids, and all the grids on the route of the sampling device can be observed { over}  certain frequencies.

\begin{figure}
		\centering
\includegraphics[width=.95\linewidth]{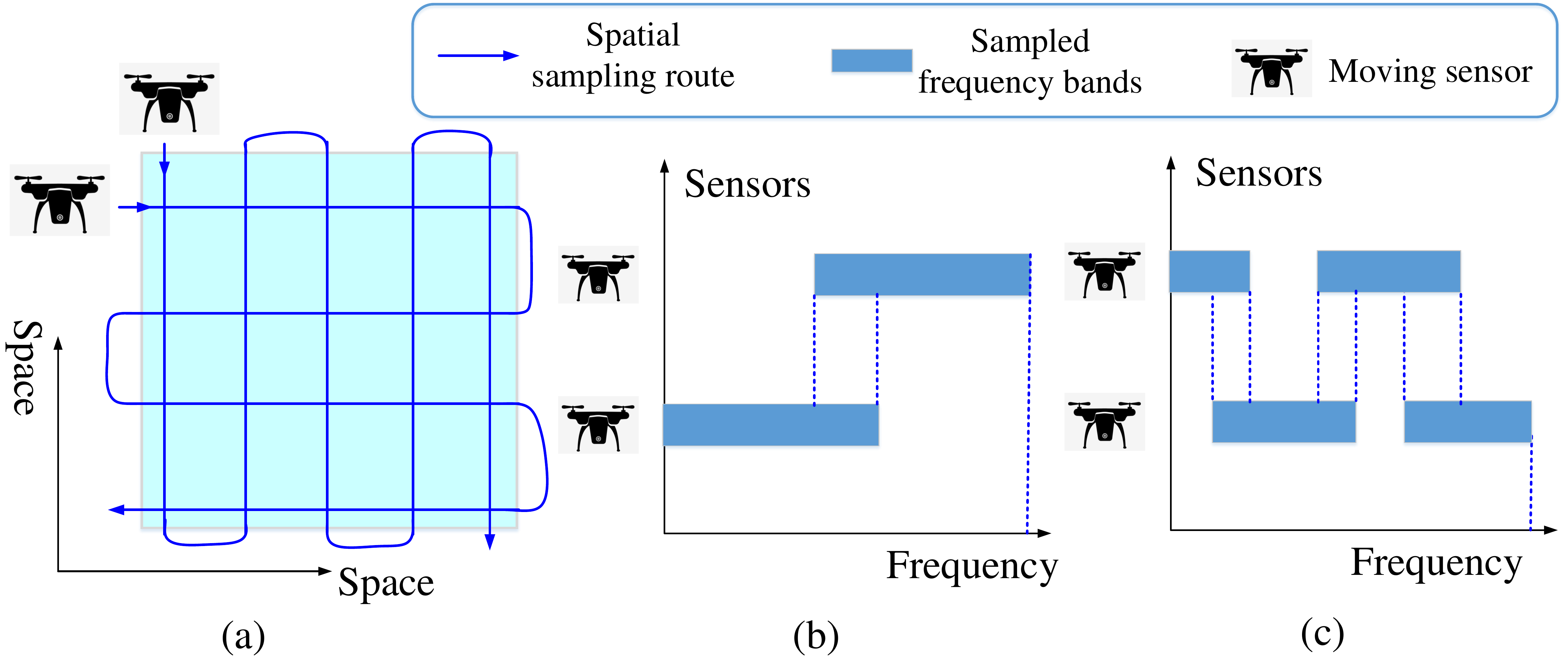}
	\caption{Using moving sensors to sample the RF tensor. (a) An example of the sampling routes using two moving sensors; (b) an example of sampled frequency bands by two sensors; (c) another example of frequency band allocation to two sensors. }
	\label{UAV}
\end{figure}

To illustrate our idea, we use a two-sensor scenario as an example---but the approach can be readily generalized to multiple sensor cases.
One possible sampling scheme is illustrated in Fig.~\ref{UAV}.
In this example, two sensors are employed---they are responsible for horizontal route and the vertical route in Fig.~\ref{UAV}, respectively.
In addition, the two sensors cover different (but overlapping) frequency bands.

Assume that the moving sensors collect power spectral measurements at every $(i,j)$ grid on their routes over the assigned frequency bands.
After the moving sensors completing the designed tour, two sets of samples of the RF tensor are collected.
To be { precise}, let ${\cal S}_1\subset\{1, \cdots, I\}$ and ${\cal S}_2\subset\{1, \cdots, J\}$ contain the observed indices along the two spatial dimensions by the two sensors, respectively. Similarly, let ${\cal S}_3,\cS_4\subset\{1, \cdots, K\}$ be the observed indices along the frequency dimensions by the two sensors, respectively.
Then, two subtensors $\tX^{(1)}=\tX(\cS_1,:,\cS_3)$ and $\tX^{(2)}=\tX(:,\cS_2,\cS_4)$, with $\vert{\cal S}_1\vert=M$, $\vert{\cal S}_2\vert=N$, $\vert\cS_3\vert=K_1$ and  $\vert\cS_4\vert=K_2$ which satisfy $\cS_3\cup\cS_4=[K]=\{1,\ldots,K\}$.
Define row-selection { matrix} $\MP = \I_I ({\cal S}_1, :)$, which consists of the rows indexed by ${\cal S}_1$ of an $I\times I$ identity matrix. Similarly, let $\Q = \I_J ({\cal S}_2, :)$, $\R_1=\I_K(\cS_3,:)$ and $\R_2=\I_K(\cS_4,:)$. The sampled measurements by sensor 1 and sensor 2 can be represented as follows:
\begin{subequations}\label{eq:Xslab}
	\begin{align}
	\tX^{(1)} &=\tX \times_1 \bm P \times_3 \bm R_1= \tX(\cS_1,:,\cS_3),\\
	\tX^{(2)} &=\tX \times_2 \bm Q \times_3 \bm R_2=\tX(:,\cS_2,\cS_4).
	\end{align}
\end{subequations}


\subsection{Coupled Tensor Decomposition-Based Formulation}
To proceed, recall our postulate that $\S_r= \A_r\B_r^\top$ where $\A_r\in\mathbb{R}^{{ I\times L}}$ and $\B_r\in\mathbb{R}^{{ J\times L}}$.
Hence, our sampled data can be written as:
\begin{subequations}\label{Xsub}
	\begin{align}
	\tX^{(1)}& = \sum_{r=1}^R ({(\bm P\A_r)}\B_r^\top)\circ (\R_1\vc_r),\\
	 \tX^{(2)} &= \sum_{r=1}^R (\A_r(\B_r^\top\Q^\top))\circ (\R_2\vc_r) .
\end{align}
\end{subequations}
It is not hard to notice that  $\tX^{(1)} $ and $\tX^{(2)} $ are two tensors that still follow the {\sf LL1} model.
More importantly, the latent factors of the original tensor, i.e., the matrices { $\A_r$, $\B_r$ for $r=1,\ldots,R$} and $\C$, are captured in the two observations.
Intuitively, if both  $\tX^{(1)}$ and $\tX^{(2)}$ admit unique {\sf LL1} decompositions, then,
one may  estimate $\A_r$ from $\tX^{(2)}$ and estimate $\B_r$ from $\tX^{(1)}$ respectively, and then `assemble' the information to get $\S_r$.

The idea is seemingly appealing, but has several significant hurdles. 
First, recall that the {\sf LL1} decomposition does not give exact $\A_r$, $\B_r$, but only the range spaces of them \cite{de2008decompositions3}.
To be specific, by decomposing $\tX^{(1)}$, one will have $\widehat{\B}_r=\B_r\bm F_r$ if the BTD problem is optimally solved, where $\bm F_r\in\mathbb{R}^{ L\times L}$ is nonsingular. Similar, from $\tX^{(2)}$, one will have $\widehat{\A}_r=\A_r\bm G_r$. In general, $\bm G_r \neq\bm F_r^{-\top}$, and thus $\S_r \neq \widehat{\A}_r\widehat{\B}_r^{\top}$, which means that the reconstruction fails.
Second, if one of $\tX^{(1)}$ or $\tX^{(2)}$ does not admit a unique {\sf LL1} decomposition, then individually decomposing the two tensors will not lead to sensible results.

To handle these challenges, we propose the following coupled factorization criterion:
\vspace{5pt}
\begin{mdframed}
\begin{align}\label{eq:coupledBTD}
\minimize_{\{\A_r,\B_r\},\C} &~\left\|  \tX^{(1)}- \sum_{r=1}^R ((\bm P\A_r)\B_r^\top)\circ (\R_1\vc_r)\right\|_F^2 \\
& +\left\| \tX^{(2)}- \sum_{r=1}^R (\A_r(\B_r^\top\Q^\top))\circ (\R_2\vc_r) \right\|_F^2.\nonumber
\end{align}
\end{mdframed}
\vspace{5pt}
The idea is to use \emph{shared} parametrization to enforce that the two `marginalized' tensors $\tX^{(1)}$ and $\tX^{(2)}$ share the same $\A_r$, $\B_r$ and $\C$.
We show that the following theorem holds:
\begin{Theorem}\label{CoupledBTD}
Assume that (\ref{eq:Xslab}) and (\ref{Xsub}) hold, and that $\A_r$, $\B_r$ for all $r$ and $\C$ are drawn from any joint absolutely continuous distributions. Also assume that $\vert\cS\vert\ge R$, where $\cS=\cS_3\cap\cS_4$.
Suppose that one of the following conditions holds
\begin{itemize}
  \item[(1)] $M\ge 2L$, $J\ge LR$, and $\min(\lfloor M/L \rfloor, R) + \min(\lfloor J/L\rfloor ,R )\ge R+2 $;
  \item[(2)] $N\ge 2L$, $I\ge LR$ , and $\min(\lfloor N/L \rfloor, R) + \min(\lfloor I/L\rfloor ,R )\ge R+2 $.
\end{itemize}
Then, solving Problem~\eqref{eq:coupledBTD} identifies $\{\S_r,\vc_r\}$ for $r=1\ldots,R$ up to scaling and permutation ambiguities, almost surely.
\end{Theorem}
The proof is relegated to Appendix~\ref{app_t3}.
The proof is reminiscent of coupled tensor decomposition under the CPD model \cite{kanatsoulis2018hyperspectral,kanatsoulis2019regular}---whereas the proof of Theorem~\ref{CoupledBTD} is more challenging since $\A_r$ and $\B_r$ are not identifiable (as opposed to the case under the CPD model).
Theorem~\ref{CoupledBTD} says that if the SLFs have low rank and the number of emitters { is} not too large, then, the proposed sampling strategy and the formulation guarantee recovering $\S_r$ and $\vc_r$ for $r=1,\ldots, R$, given $M$ and $N$ (i.e., the numbers of the partially sampled horizontal and vertical slabs, respectively) exceeding a certain threshold.

\begin{Remark}{\rm
	We would like to mention some interesting implications of Theorem~\ref{CoupledBTD}. First, the sensors are only required to collect data from { a} certain number of slabs, but it does not matter which slabs should be used. In particular, the sampled slabs do not necessarily need to be equi-spaced.
	Second, the sensing pattern is completely regular, without involving randomness. These properties entail the system designers flexibility to circumvent practical issues.}
\end{Remark}

\begin{Remark}{\rm
	Theorem~\ref{CoupledBTD} addresses identifiability of the two-sensor case that we used as an illustrative example. In practice, many more sampling patterns involving more sensors can be shown to have similar results. Nevertheless, The goal of this work is offer proof-of-concept, rather than exhausting all feasible scenarios.}
\end{Remark}

\subsection{The Static Sensor Case: Sensing Scattered Grids}
Moving sensors have the advantage of high mobility and flexibility.
However, moving sensors are not always available.
In this subsection, we consider scenarios where sensors are deployed in the geographical area in advance and sense at their respective locations simultaneously. Such a setting is more conventional \cite{ureten2012comparison,hamid2017non,teganya2018localization,jayawickrama2013improved}.

%

While in theory the sampling patterns discussed in the previous subsection (cf. Fig.~\ref{UAV}) can also be applied using static sensors, some times one may also hope to place the sensors over the region in a scattered way (rather than placing the sensors in consecutive grids like in Fig.~\ref{UAV}).
This kind of sensor deployment is related to a concept called {\it fiber sampling} \cite{kanatsoulis2019regular,sorensen2017fiber}.
A fiber of the tensor $\tX$ refers to $\tX(i,j,:)$; see Fig. \ref{fig:modes}. In our context, a fiber is a power spectrum of the received signal measured at the geographical location $(i,j)$. We assume that the sensor at location $(i,j)$ collects the spectral information of bands indexed by ${\cal K}\subseteq \{1,\ldots,K\}$.  Then, the collected data is denoted as $\tX(i,j,\CK)$---which can be the whole fiber or a part of the fiber $\tX(i,j,:)$.

\subsubsection{Systematic Sensor Deployment}
To proceed, let us consider a scenario where the locations of sensors are under control---i.e., system designers can design the deployment strategy of the sensors.
Suppose that we have $D$ groups of sensors.
We assume that group $d$ collects data from bands indexed by ${\cal K}^{(d)}$.
In addition, assume that the locations of group $d$ sensors are indexed by all combinations possible $(i_d,j_d)$ such that $i_d\in{\cal I}^{(d)}$ and $j_d\in{\cal J}^{(d)}$.
This way, the sampled data from group $d$ forms a subtensor  $\tX\dd=\tX(\CI\dd,\CJ\dd,\CK\dd)$. 
%
Since $\tX$ admits a rank-{ $(L,L,1)$} BTD, the sampled subtensor $\tX\dd$ can be written as:
\begin{equation}\label{Xd}
\tX\dd=\sum_{r=1}^R(\A_r(\CI\dd,:)\B_r(\CJ\dd,:)^\top)\circ \vc_r(\CK\dd).
\end{equation}

To estimate $\C$ and $\S_r$, we propose the following criterion:
\begin{mdframed}
	\begin{align}\label{eq:coupledBTD_fiber}
	\minimize_{\{\A_r,\B_r\},\C} &~\sum_{d=1}^{D}\left\| \tX\dd-\sum_{r=1}^R\S_r({\cal I}^{(d)},{\cal J}^{(d)})\circ \vc_r(\CK\dd)\right\|_F^2,
	\end{align}
\end{mdframed}
\vspace{7pt}
where we have
$ \S_r({\cal I}^{(d)},{\cal J}^{(d)})= \A_r(\CI\dd,:)\B_r(\CJ\dd,:)^\top.$
In terms of identifiability of the SLFs and PSDs, we show the following theorem:

\begin{Theorem}\label{MultiTensor}
Assume that (\ref{Xd}) holds.
Also assume that $\A_r$, $\B_r$ for all $r$ and $\C$ are drawn from certain joint absolutely continuous distributions.
Suppose that index sets $\CI\dd,\CJ\dd,\CK\dd$ satisfy
\begin{equation}\label{eq:union}
\begin{aligned}
&\cup_{d=1}^D\CI\dd =[I],~
\cup_{d=1}^D\CJ\dd =[J],~
\cup_{d=1}^D\CK\dd =[K],
\end{aligned}
\end{equation}
$
\vert\CI\dd\vert\ge L, \vert\CJ\dd\vert\ge L,\vert\CK\dd\vert\ge R
$, and that
\[
\min\left(\bigg\lfloor\frac{\vert\CI\dd\vert}{L}\bigg\rfloor,R\right) + \min\left(\bigg\lfloor\frac{\vert\CJ\dd\vert}{L}\bigg\rfloor,R\right)\ge R+2.
\]
Assume that there exists a permutation of set $[D]$, denoted as $\{i_1,\cdots,i_D\}$, satisfying
$
\max(\vert\CI\id\cap\CI\idp\vert,\vert \CJ\id\cap\CJ\idp\vert)\ge L
$
and
$
\vert\CK\id\cap\CK\idp\vert\ge 2
$
for $d=1,\cdots,D-1$. Then, $\{\S_r,\vc_r\}_{r=1}^R$ can be identified via solving Problem~\eqref{eq:coupledBTD_fiber} up to scaling and permutation ambiguities almost surely.
\end{Theorem}

\begin{figure}
		\centering
	\includegraphics[width=.85\linewidth]{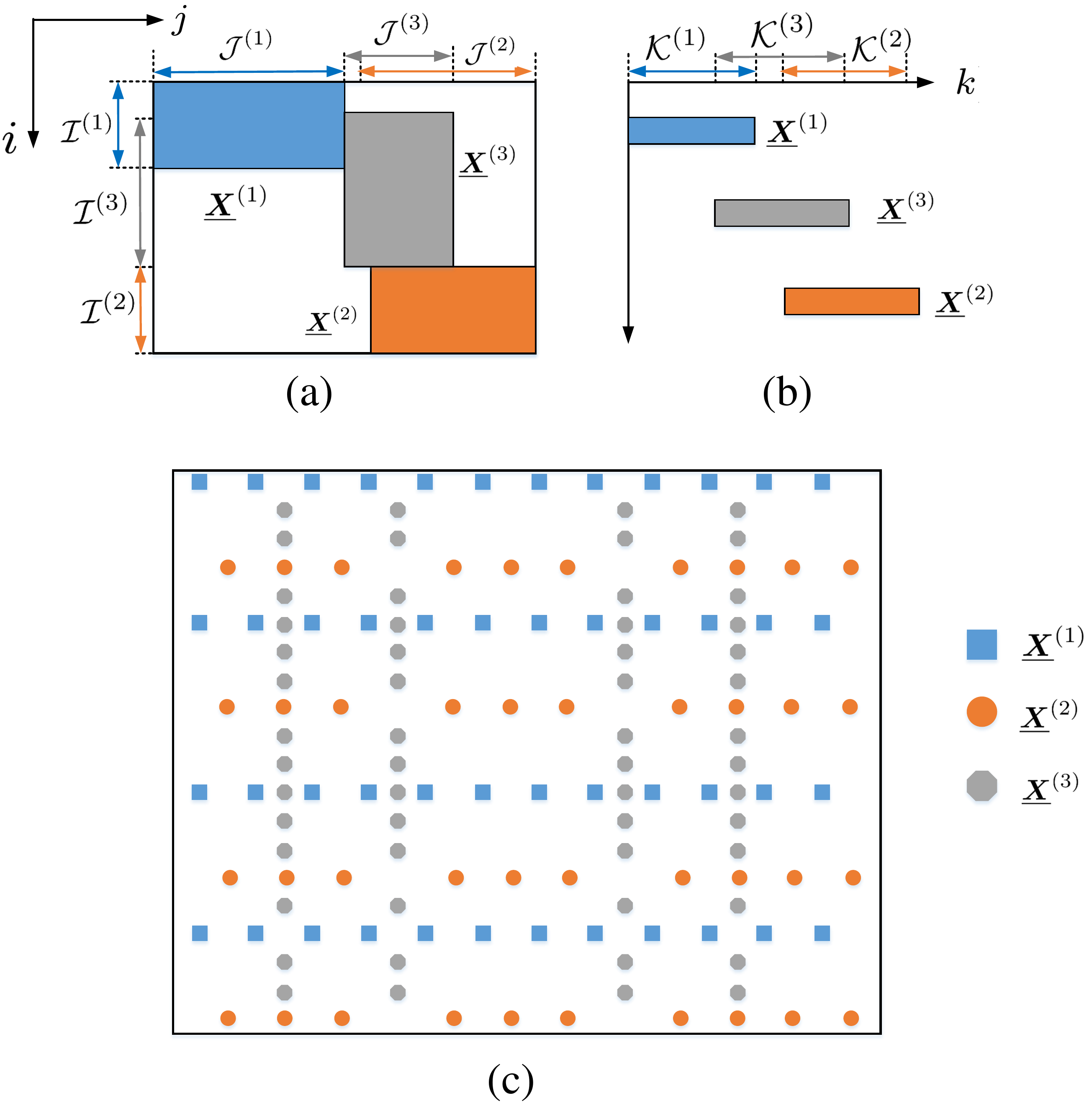}
	\caption{An example of $D=3$ subtensors that satisfy the conditions of Theorem~\ref{MultiTensor}---$i_1=1,i_2=3,i_3=2$ and $\vert\CI^{(i_1)}\cap\CI^{(i_2)}\vert\ge L$, $\vert\CJ^{(i_2)}\cap\CJ^{(i_3)}\vert\ge L$,$\vert\CK^{(i_1)}\cap\CK^{(i_2)}\vert\ge 2$, $\vert\CK^{(i_2)}\cap\CK^{(i_3)}\vert\ge 2$. (a) The overlapped index sets of the first and second dimensions of $\tX^{(1)},\tX^{(2)},\tX^{(3)}$. (b) The overlapped index sets of the third dimension of $\tX^{(1)},\tX^{(2)},\tX^{(3)}$. (c) A `scattered' version by permuting the spatial sampling pattern in (a).}
	\label{MultiTensorFig}
\end{figure}

The proof is relegated to Appendix \ref{app_t4}.
Fig. \ref{MultiTensorFig} shows an example satisfying the condition in Theorem~\ref{MultiTensor} where $D=3$ groups of sensors are deployed. In particular, Fig. \ref{MultiTensorFig} (c) shows a deployment pattern where the sensors are scattered rather than being placed in consecutive grids. 

One caveat is that Theorem~\ref{MultiTensor} requires all { $\tX^{(d)}$ for $d=1,\ldots,D$} being identifiable {\sf LL1} tensors. However, the following theorem shows that, if there is one group of sensors that collect samples from all the frequency bands, and the corresponding subtensor is identifiable, the whole tensor is identifiable:
\begin{Theorem}\label{MultiTensor-C}
Assume that (\ref{Xd}) holds.
Also assume that $\A_r$, $\B_r$ for all $r$ and $\C$ are drawn from certain joint absolutely continuous distributions.
Suppose that the index sets $\CI\dd,\CJ\dd,\CK\dd$ satisfy
$
\cup_{d=1}^D\CI\dd =[I],
 \cup_{d=1}^D\CJ\dd =[J].
$
and
$
\vert\CI\dd\vert\ge L, \vert\CJ\dd\vert\ge L
$
for $d=1,\cdots,D$. Assume that there exists $d_0\in[D]$ that satisfies $\CK\dz=[K]$. In addition, assume that $K\ge R$ and
\[
\min\left(\bigg\lfloor\frac{\vert\CI\dz\vert}{L}\bigg\rfloor,R\right) + \min\left(\bigg\lfloor\frac{\vert\CJ\dz\vert}{L}\bigg\rfloor,R\right)\ge R+2,
\]
and that there exists a permutation of set $[D]$, denoted as $\{i_1,\cdots,i_D\}$, satisfying
$
\max(\vert\CI\id\cap\CI\idp\vert,\vert \CJ\id\cap\CJ\idp\vert)\ge L
$
for $d=1,\cdots,D-1$.
Then, $\{\S_r,\vc_r\}_{r=1}^R$ can be identified via solving Problem~\eqref{eq:coupledBTD_fiber} up to scaling and permutation ambiguities almost surely.
\end{Theorem}

The proof is relegated to Appendix \ref{app_t5}. One can see that compared to Theorem~\ref{MultiTensor}, the above theorem only needs one of $\tX^{(d)}$, namely $\tX^{(d_0)}$, to be identifiable, and the { remaining} subtensors do not need to be identifiable and are only required to overlap at least $L$ indices in the first or second { dimension}. However, sensors in group $d_0$ are required to sample all the frequency bands of interest under Theorem~\ref{MultiTensor-C}, while none of the groups under Theorem~\ref{MultiTensor} are required to do so---but they need to cover larger space in order to recover the same { tensor} $\tX$. Apparently, these two theorems present an interesting spectral-spatial sampling trade-off.

\begin{Remark}{\rm
	The work in \cite{sorensen2017fiber} considered a similar setting as that in Theorems~\ref{MultiTensor}-\ref{MultiTensor-C} and showed identifiability of a coupled {\sf LL1} decomposition criterion. Notably, the results in \cite{sorensen2017fiber} can also be applied to our case. The key difference is that the work in \cite{sorensen2017fiber} assumes full observation on the third dimension of the tensor, which means that every sensor observes all the frequency bands of interest---which is not assumed in Theorems~\ref{MultiTensor}-\ref{MultiTensor-C}. The work in \cite{kanatsoulis2019regular} also considered fiber sampling and tensor recovery, but under the CPD model instead of the ${\sf LL1}$ model.

{ On a higher level, Theorems 1-5 in this work can be understood as coupled {\sf LL1} decomposition for downsampled tensor recovery with regular/systematical sampling strategies---which share insights with tensor recovery from coupled CPD models \cite{kanatsoulis2018hyperspectral,kanatsoulis2020tensor}. The new challenge for establishing recoverability under the the coupled {\sf LL1} framework lies in the non-identifiability of $\A_r$ and $\B_r$, which needs extra care to handle in the recoverability proofs.}
}
\end{Remark}

%
%



\subsubsection{Random Sensor Deployment}
Many existing radio map recovery methods considered random sampling over the space.
As a side contribution, we also show that under random sampling radio map disaggregation is possible.
Let us consider a formulation when randomly deployed sensors are used:
\vspace{5pt}
\begin{mdframed}
\begin{equation}
\minimize_{\A,\B,\C} \; \left\Vert\tW\ast \left(\tX-\sum_{r=1}^R (\A_r\B_r^\top)\circ \vc_r \right) \right\Vert_F^2, \label{opt_rndfib}
\end{equation}
\end{mdframed}
where $\ast$ denotes the Hadamard product (also known as entrywise product) and $\tW$ is a tensor such that $ \tW(i,j,k)=1$ if $(i,j,k)\in\bm \Omega$, and  $ \tW(i,j,k)=0$ otherwise.
%
%
In terms of identifiability, it can be seen that

\begin{Prop}\label{FibSampTheorem}
Under the exact {\sf LL1} model of $\tX$,
assume that, w.l.o.g., $I\leq J$, $J> (LR+1)(I-LR)$ and $LR\le \frac I6$. Also assume that $\{\A_r,\B_r\}_r$ and $\C$ are drawn from any joint continuous distribution.
Suppose that noise is absent.
Given $0<\epsilon\le 1$, if $q$ entries per column of the frontal slabs $\tX(:,:,k),k=1,\cdots,K$ are observed uniformly at random, where
$
q\ge \max\left\{12\log\left(\frac{I}{\epsilon}+1\right),2LR\right\},
$
then solving Problem~\eqref{opt_rndfib} recovers $\{\S_r,\vc_r\}$ for $r=1\ldots,R$ with a probability at least $1-\epsilon$.
\end{Prop}

The proof is presented in Appendix \ref{app_t6} in the supplementary { material} and is straightforward.
In a nutshell, if the matrix rank of each slab $\tX(:,:,k)$ is low enough, then the slabs are identifiable---via \textit{matrix completion}.
Then, the SLFs and PSDs are identifiable by classic {\sf LL1} model identifiability results. Following this insight, and with careful derivations, the above theorem can be shown.

	We would like to remark that Proposition~\ref{FibSampTheorem} merely states that the SLFs and PSDs of interest can be identified under random sensor deployment. However, it does not demonstrate advantages in theory for using the {\sf LL1} model-based formulation as in \eqref{opt_rndfib}, as opposed to performing slab-by-slab matrix completion and then factoring the completed tensor into block terms. Investigating these aspects is quite intriguing, but may go beyond the scope of this work. We defer this to a potential future work. Nonetheless, in practice, using the formulation in \eqref{opt_rndfib} may be beneficial for a number of reasons. For example, explicitly using prior knowledge (the underlying {\sf LL1} model in this case) is known to be helpful in fending against noise.
	Also using an all-at-once optimization criterion for estimating the PSDs and SLFs is less possible to suffer from error propagation, compared to slab-by-slab matrix completion and then applying ${\sf LL1}$ decomposition, as we will see shortly in the simulations.
%
%
%


\section{Algorithms and Practical Implementation}
In this section, we propose algorithms for handling the formulated factorization criteria in the previous section.

\subsection{Algorithm for Solving \eqref{eq:coupledBTD}}
Let us denote the objective function of \eqref{eq:coupledBTD} as $f(\A,\B,\C)$.
In practice, instead of trying to minimize $f(\A,\B,\C)$, we use the following regularized version
\begin{align} \label{Jval_form}
\minimize_{\{\A_r,\B_r\},\C} ~\underbrace{f(\A,\B,\C) + \lambda_1\|\A\|_F^2+\lambda_2 \Vert\B\Vert_F^2 + \lambda_3\Vert\C\Vert_F^2}_{{\sf Loss}(\A,\B,\C)}.
\end{align}
{ The regularization terms are added to alleviate the so-called scaling/counter-scaling effect. Specifically, $\alpha(\A_r\B_r^\T)  \circ (1/\alpha)\C(:,r)$ may happen without changing the model fitting error, where $\alpha$ could be arbitrarily large, thereby affecting numerical stability. Adding regularization can prevent such pathological cases from happening; see discussions in \cite{fu2015joint}.}


We propose to employ a \textit{block coordinate descent} (BCD) scheme for handling Problem~\eqref{Jval_form}. To be specific, $\A,\B,\C$  are updated as follows:
\begin{subequations}\label{eq:BCD}
	\begin{align}
	 \A^{t+1} &\leftarrow \arg\min {\sf Loss}(\A,\B^t,\C^t) \label{Aeq:BCD}\\
	 \B^{t+1} &\leftarrow \arg\min {\sf Loss}(\A^{t+1},\B,\C^t)\label{Beq:BCD}\\
	  \C^{t+1} &\leftarrow \arg\min {\sf Loss}(\A^{t+1},\B^{t+1},\C)\label{Ceq:BCD},
	\end{align}
\end{subequations}
where superscript $t$ is the iteration index.
Note that all the subproblems above are unconstrained least squares problems and thus can be readily solved; see details in Appendix~\ref{app_BC} in the supplementary { material}.

%

As a standard exact BCD algorithm, the algorithm in \eqref{eq:BCD} enjoys convergence properties as discussed in \cite{razaviyayn2013unified,bertsekas1997nonlinear}.
Specifically, every limit point of the produced solution sequence is a stationary point of \eqref{Jval_form}, when $\lambda_n>0$ for $n=1,2,3$---since the subproblems are always strictly convex.

\subsection{Algorithm for Solving Problems \eqref{opt_rndfib}}
The basic idea for handling \eqref{opt_rndfib} is again BCD as in \eqref{eq:BCD}. Nevertheless, since the sampling pattern is quite different, the subproblems are solved in different ways.
To be specific, defining $\tY=\tW\ast\tX$, let $\Y_n$ and  $\W_n$ be mode-$n$ unfoldings of $\tY$ and $\tW$, respectively. The update rules for the regularized version of Problem ~\eqref{opt_rndfib} are as follows:
\begin{align*}
\A&\leftarrow\arg\min_{\A} \Vert\Y_1-\W_1\ast(\C\odot_p\B)\A^\top\Vert_F^2+ \lambda_1\Vert\A\Vert_F^2 \\
\B&\leftarrow\arg\min_{\B} \Vert\Y_2-\W_2\ast(\C\odot_p\A)\B^\top\Vert_F^2 + \lambda_2\Vert\B\Vert_F^2\\
\C&\leftarrow\arg\min_{\C}
\Vert\Y_3-\W_3\ast\S\C^\top\Vert_F^2+ \lambda_3\Vert\C\Vert_F^2,
\end{align*}
where $\S=[\tvec(\A_1\B_1^\top),\cdots,\tvec(\A_R\B_R^\top)]$.
Each subproblem boils down to a set of unconstrained least squares problems---see details in Appendix \ref{app_ABCfiber} in the supplementary { material}.

\begin{Remark}{\rm
{ Problem~\eqref{eq:coupledBTD_fiber} can be handled by the same algorithm}, with a slightly different definition of $\underline{\bm W}$---to take into consideration that group $d$ and group $d'$ are overlapped in space. To be specific, if $(i,j,k)$ is sensed by $P$ groups simultaneously, then we have
$    \underline{\bm W}(i,j,k)=\sqrt{P}. $
	Under this definition of $\underline{\bm W}$, { Problem~\eqref{eq:coupledBTD_fiber} can be re-formulated in the form of \eqref{opt_rndfib}, thereby  the same algorithm can be used.}}
\end{Remark}

\subsection{Performance Enhancement}
Coupled tensor decomposition is nonconvex and hard to compute, even without noise.
In practice, noise can make the estimation for the SLFs and PSDs even harder.
In this subsection, we propose a series pragmatic engineering tricks, e.g., suitable initialization and post-processing, to help enhance performance of the propose approach.
\subsubsection{Estimating $\S$}
Our identifiability theorems assume that { $\S_r$s} are exact low-rank matrices.
However, in practice, the SLFs are approximately low-rank.
Consequently, we notice that the estimation accuracy of the algorithm-output $\C$ is usually (much) better than that of $\widehat{\S}_r=\widehat{\A}_r\widehat{\B}_r^\T$. It is of interest to exploit this empirical observation.

To this end, note that in (\ref{Xunfold}b), $\X_3(\ell,:)=\S(\ell,:)\C^\T$, where $\S=[\tvec(\S_1),\cdots,\tvec(\S_R)]$. If $\C$ is accurately estimated and $\C$ has full column rank, then,
$  \widehat{\S}(\ell,:)= \X_3(\ell,:) (\widehat{\C}^\T)^\dagger$
is a fairly accurate estimate for $\S(\ell,:)$. Note that $\S(\ell,r)=\S_r(i,j)$ for $\ell = (i-1)I + j$---that is, there are a lot of elements of $\S_r$ can be accurate estimated. The remaining task is to utilize such information to interpolate the other elements. This, as a 2-D interpolation problem, is very well-studied in the literature \cite{li2011review,ikechukwu2017accuracy}. Since $\S_r$ is an SLF, it is smooth over space and easy to interpolate. In this work, we adopt the widely used spline interpolators, the \textit{thin-plate splines} (TPS) \cite{wahba1990spline}. After estimating $\widehat{\S}$, the estimate of $\widehat{\X}$ can be obtained from its unfolding $\widehat{\X}_3=\widehat{\S}\widehat{\C}^\T$
\subsubsection{Initialization}
There are many ways for offering initializations to the BCD algorithms. For example, to initialize the algorithm for handling \eqref{eq:coupledBTD}, one can decompose the `larger' (and { thus  more likely to} admit a unique {\sf LL1} decomposition) subtensor using {\sf LL1} and extract partial information of $\A_r$, $\B_r$ and $\C$. Then, the full latent factors can be estimated following the procedures of initialization techniques for coupled CPD as in \cite{kanatsoulis2018hyperspectral,kanatsoulis2019regular}. We omit the details due to space limitations.

\begin{Remark}
	{\rm Another possible performance-enhancement strategy is to incorporate prior information on $\bm c_r$ and $\S_r$, e.g., nonnegativity (per their physical meaning) and spatial/spectral smoothness. In terms of algorithm design, adding nonnegativity constraints to the formulated problem \eqref{Jval_form} is readily within reach, which leads to solving a series of nonnegativity-constrained least squares problems for updating the latent factors. Using structural information may help fend against heavy noise or shadowing. }
\end{Remark}


\section{Simulations}\label{SecSimulation}

\begin{figure}[!htbp]
	\centering
    \includegraphics[width=0.9\linewidth]{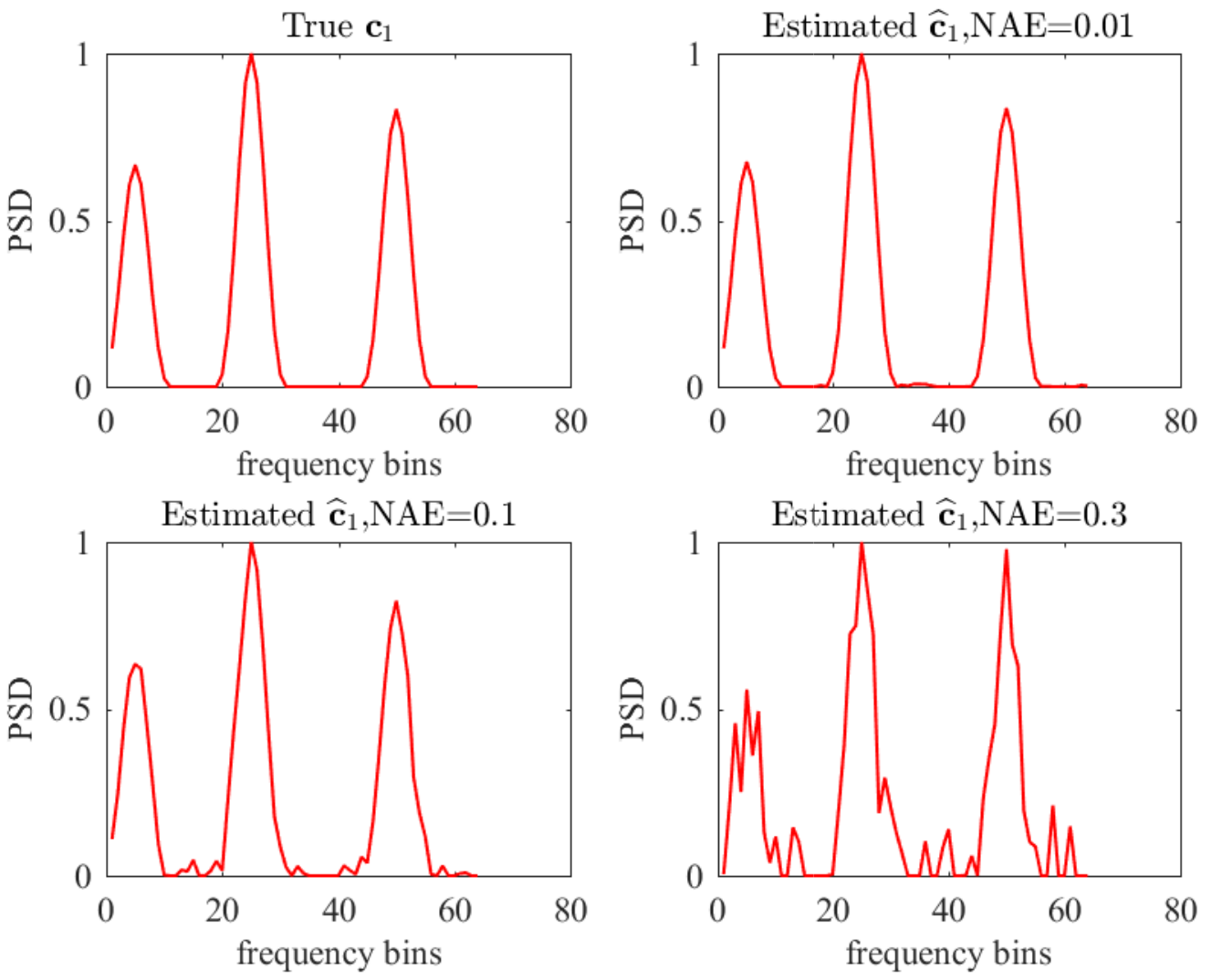}
\caption{Illustration of the ground-truth $\vc_1$ and the estimated $\widehat{\vc}_1$ corresponding to different NAEs.}
	\label{NAEC}
\end{figure}

In the simulations, the bands of interest are divided into $K$ frequency bins, where $K=64$ or $128$. We assume that there are $R$ emitters randomly deployed in a region with a size of 100 $\times$ 100 m$^2$. The region is discretized into { $101\times 101$ grids ($\{0,1,\cdots,100\}\times\{0,1,\cdots,100\}$), i.e., $I=J=101$}.  The PSDs of the emitters are generated as combinations of three randomly scaled squared sinc functions. This simulates a transmitter that occupies at most three carrier frequencies. { For} each of the emitters, the three carrier frequencies are randomly picked (cf. Fig.~\ref{NAEC}). To be more specific, the PSD of transmitter $r$, i.e., $\vc_r$,  is  generated by
$
\vc_r(k)=\sum_{i=1}^3 p_i^ra_i^r\text{sinc}^2\left(\frac{k-f_i^r}{w_i^r}\right),
$
where $p_i^r$ $(i=1,2,3)$ follows a binomial distribution with equal probability; $a_i^r$ follows a uniform distribution from $0.5$ to $2$;  $f_i^r\in\{1,\cdots,K\}$ is the center of the $i$-th square sinc function $\text{sinc}^2(\cdot)$; and $w_i^r$ follows a uniform distribution between $2$ and $4$.

Each SLF is generated following a path-loss model and the spatial correlated log-normal shadowing model \cite{goldsmith2005wireless}.
 The SLF  of emitter $r$ at location $\y=[i,j]^\T\in\bbR^2$ is generated by
$
\S_r(i,j) =\Vert \y-\z_r\Vert_2^{-\eta_r} 10^{ z_r(\y)/{10}},
$
where $\z_r\in\bbR^2$ denotes the location of emitter $r$; $\eta_r$ is the path-loss coefficient of transmitter $r$. The correlated log-normal shadowing component $z_r(\y)$ is generated as a zero-mean Gaussian random variable with autocorrelation
\begin{equation}\label{eq:Xc}
\mathbb{E}\{z_r(\y)z_r(\y')\}=\sigma_r^2\exp(-\Vert \y-\y'\Vert_2/X_c)
\end{equation}
between locations $\y$ and $\y'$, where $X_c$ is the decorrelation distance.
{ We take $\sigma_r = \sigma$ for all $r$.}
For a typical outdoor environment, $X_c$ varies from 50 to 100 \cite{goldsmith2005wireless}. We use $X_c=30$ to test the algorithms under { more severe shadowing effects}.
The parameter $\eta_r$ follows a uniform distribution between $2$ and $3$. { In the simulations, we set $\sigma=4,X_c=30$ unless specified.}

To evaluate the performance of estimated $\widehat{\C},\widehat{\S}$ and $\widehat{\tX}$, we first fix the permutation ambiguity by solving the following matching criterion:
\[
\widehat{\bpi}=\arg\min_{\bpi\in\Phi} \sum_{r=1}^R\bigg\Vert \frac{\vc_r}{\Vert\vc_r\Vert_1}-\frac{\widehat{\vc}_{\pi_r}}{\Vert\widehat{\vc}_{\pi_r}\Vert_1}\bigg\Vert_1
\]
where $\Phi$ is the set of all permutations of $\{1,\cdots,R\}$ and $\bpi=[\pi_1,\cdots,\pi_R],\widehat{\bpi}=[\widehat{\pi}_1,\cdots,\widehat{\pi}_R]$; and $\vc_r$ and $\widehat{\vc}_{\pi_r}$ are the true PSD of transmitter $r$ and its estimate, respectively.
In the sequel, we assume that the permutation ambiguities of $\widehat{\C},\widehat{\S}$ have been removed.

To  evaluate the estimation performance of $\widehat{\C}$ and $\widehat{\S}$, we employ
the \textit{ normalized absolute error} (NAE) for the estimated $\C$:
\[
\text{NAE}_{\C}=\frac 1{R} \sum_{r=1}^R \bigg\Vert\frac{\vc_r}{\Vert\vc_r\Vert_1}-\frac{\widehat{\vc}_{r}}{\Vert\widehat{\vc}_{r}\Vert_1} \bigg\Vert_1;
\]
the NAE of $\S_r$ (denoted as ${\rm NAE}_{\bm S}$) is defined in an identical way.
NAE of the estimated radio map is used for evaluating the radio map recovery performance, which is defined as follows:
\[
\text{NAE}_{\tX} = \frac{\sum_{k=1}^K\Vert \tX(:,;,k)-\widehat{ \tX}(:,;,k)\Vert_1}{\sum_{k=1}^K\Vert \tX(:,;,k)\Vert_1}.
\]
The proposed algorithms are terminated when the relative change of the loss function is less than $10^{-3}$ or when the number of iterations exceeds 100.
We run 100 Monte Carlo trials and take the median of the NAEs. { For the proposed method, we set $L$ for our method via empirical tuning---since we observe that the results are not sensitive to $L$. However, one may also employ existing $L$ estimation methods for ${\sf LL1}$ in the literature; see, e.g., \cite{zarzoso2017parameter,han2017block}. The parameter $R$ is the number of emitters and is assumed to be known or has been estimated.}

\begin{figure}[!htbp]
	\centering
    \includegraphics[width=0.9\linewidth]{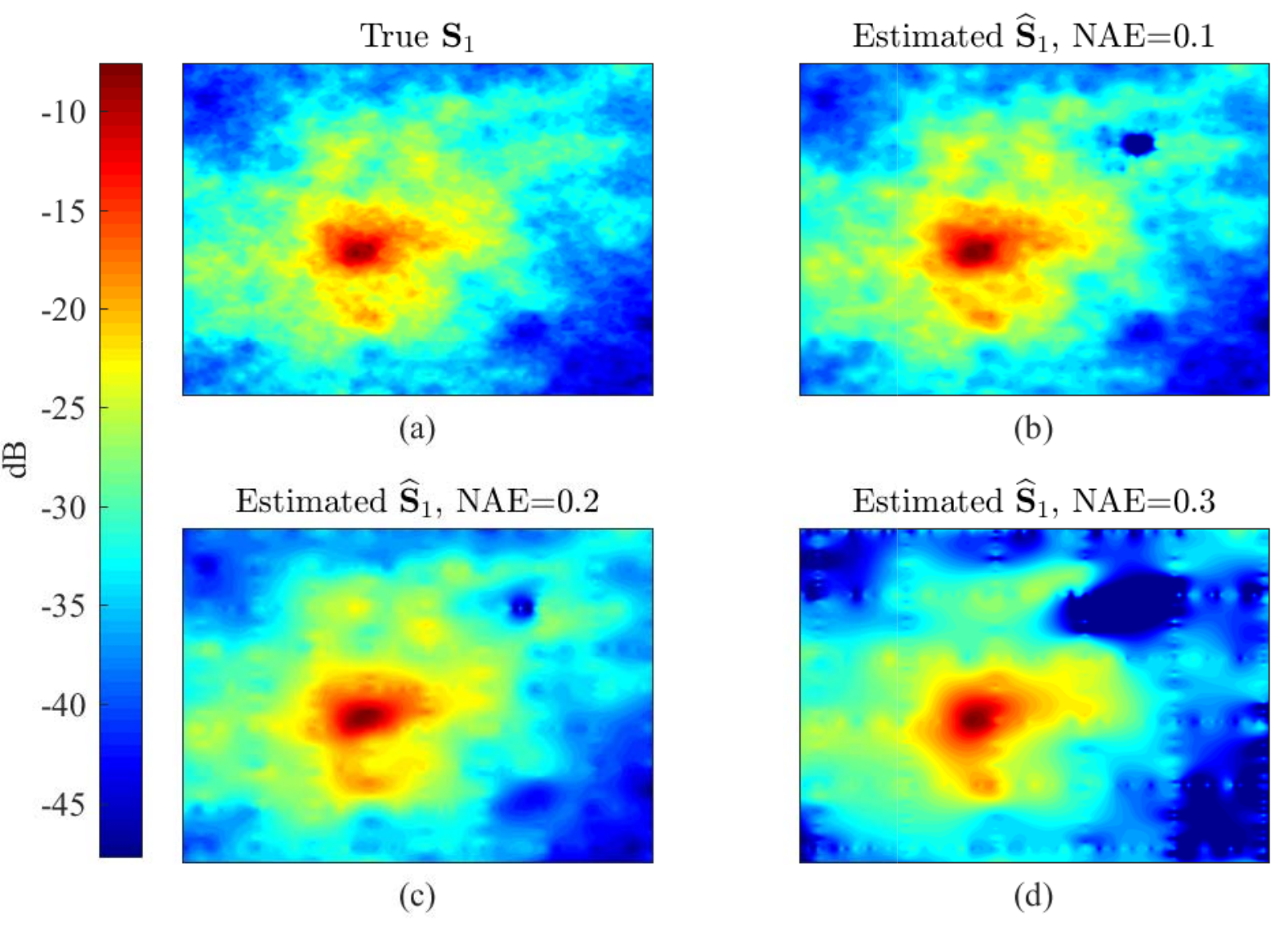}
\caption{Illustration of the ground-truth $\S_1$ and the estimated $\widehat{\S}_1$ corresponding to different NAEs.}
	\label{NAES}
\end{figure}

\begin{figure}[!htbp]
	\centering
	\includegraphics[width=0.9\linewidth]{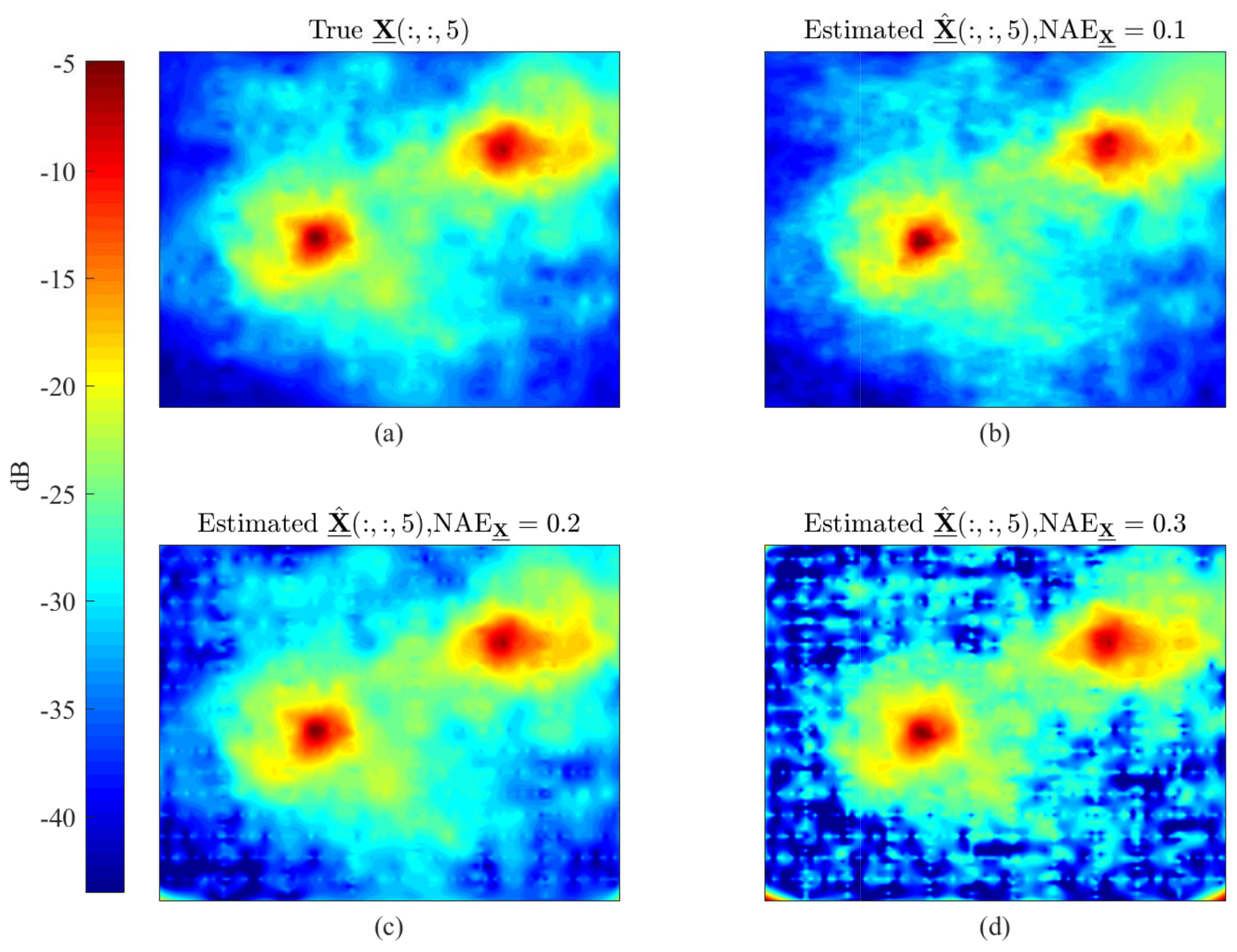}
	\caption{Illustration of the ground-truth $\tX(:,:,5)$ and the estimated $\widehat{\tX}(:,:,5)$ corresponding to different NAEs.}
	\label{NAEX}
\end{figure}

To make these metrics clearly understandable, we offer a number of references in Figs.~\ref{NAEC}-\ref{NAEX}.
For example, in Fig. \ref{NAEC}, we visualize the ground-truth PSD and the estimated PSD under NAE$_{\bm C}\approx 0.01,0.1,0.3$. When NAE$_{\bm C}\approx 0.01$, the estimated PSD is almost identical the ground-truth PSD. When ${\rm NAE}_{\C}\approx 0.3$, the estimated PSD is much worse but still readable.

\subsection{Scenario 1: Moving Sensors and Slab Sampling}

In this subsection, we consider the case where we observe two subtensors $\tX^{(1)}=\tX(\cS_1,:,:)$ and $\tX^{(2)}=\tX(:\cS_2,:)$, with $\vert{\cal S}_1\vert=M$, $\vert{\cal S}_2\vert=N$.
We use the formulation in Eq.~\eqref{eq:coupledBTD} as the disaggregation criterion.
The regularization parameters $\lambda_n$ is set to $\lambda_n=10^{-2}$ for $n=1,2,3$.

\begin{figure}[!htbp]
	\centering
	\includegraphics[width=0.9\linewidth]{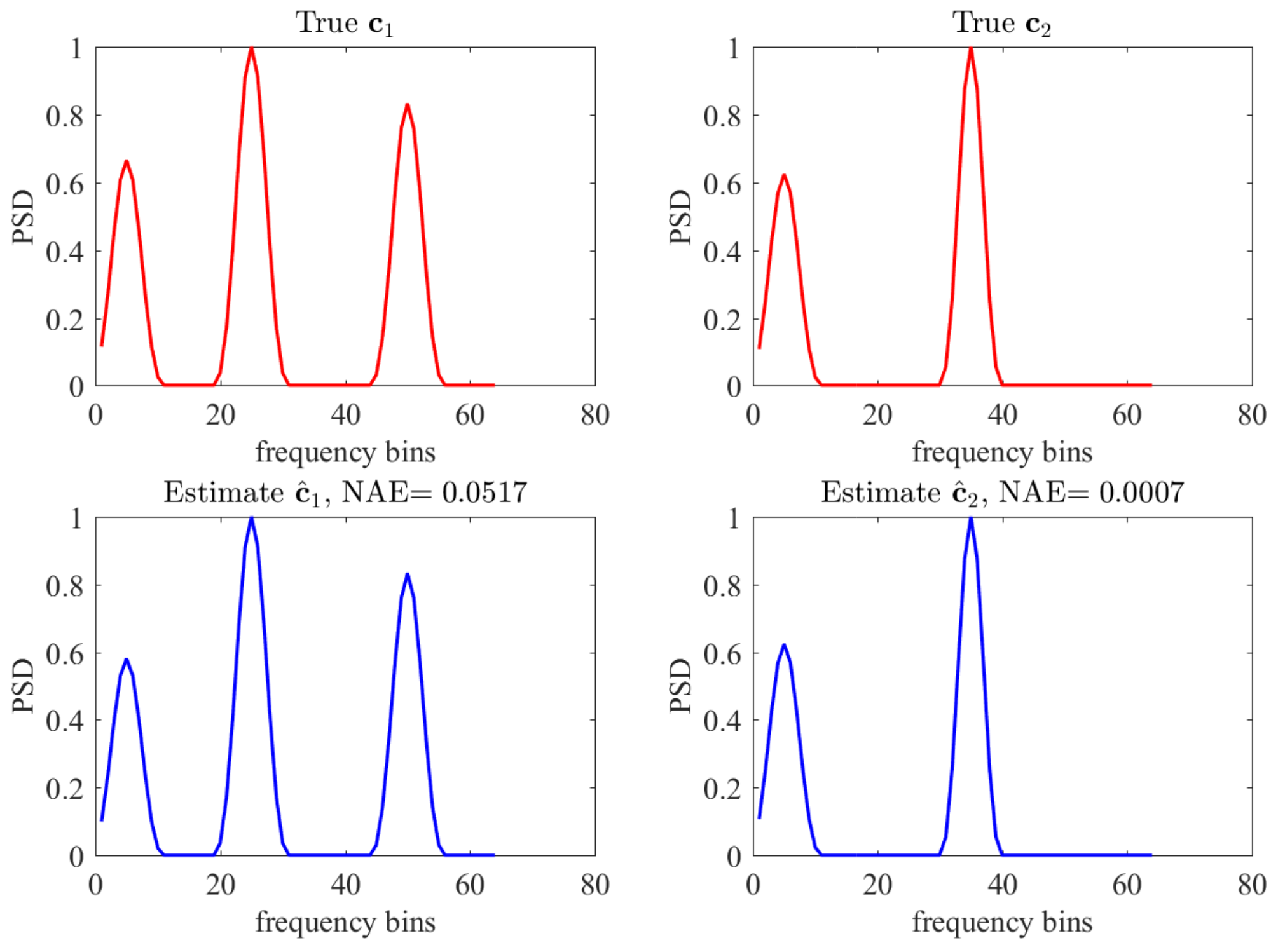}
	\caption{Ground-truth (top) and estimated (bottom) PSDs of two sources.}
	\vspace{-.25cm}
	\label{PSD}
\end{figure}

\begin{figure}[!htbp]
	\centering
	\includegraphics[width=0.90\linewidth]{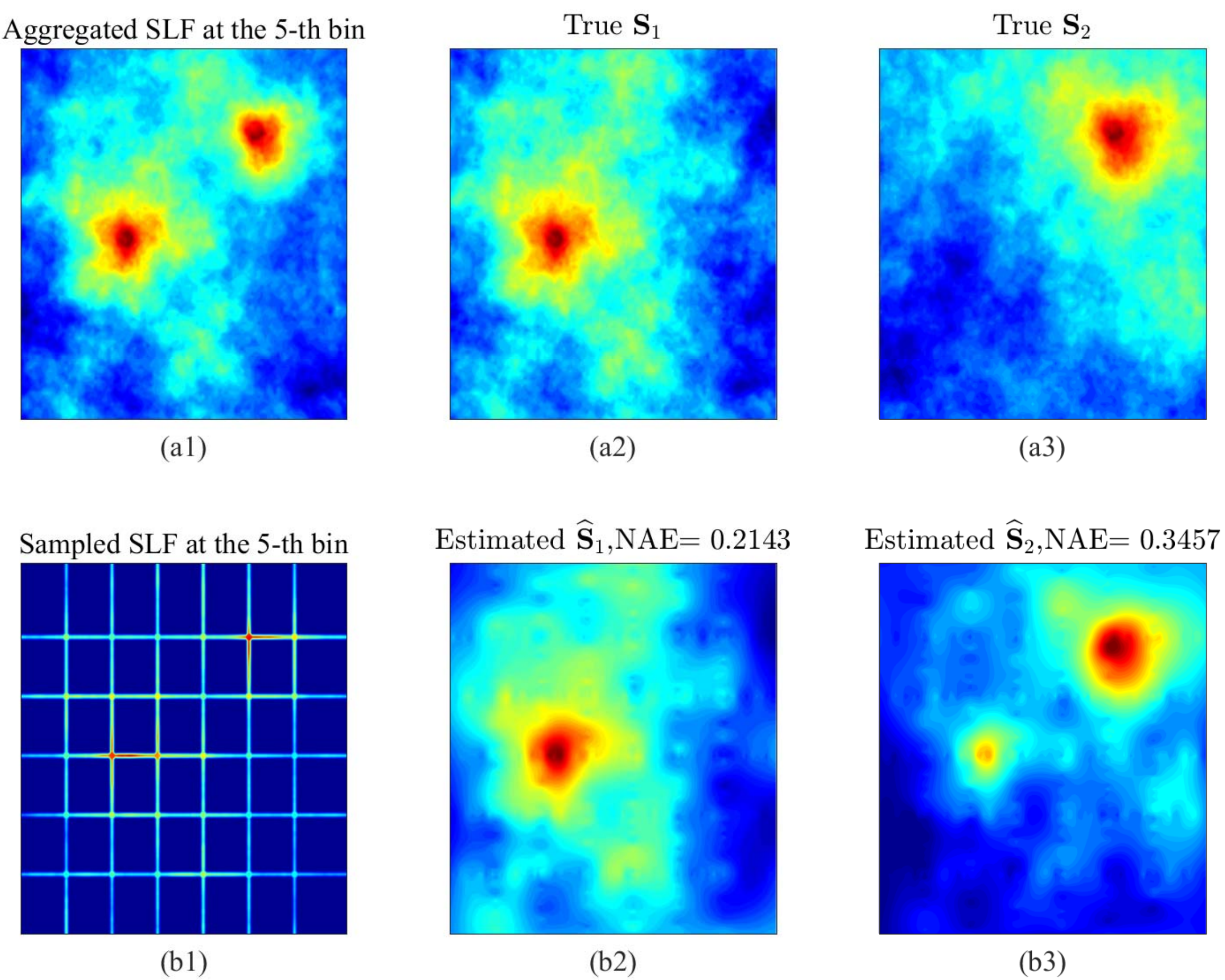}
	\caption{Ground truth: (a1) The aggregated SLF at the 5-th frequency bin; (a2) - (a3) The ground-truth SLFs. (b1) {$M=5$ rows and $N=6$} columns are sampled from $\tX$ (the unsampled entries was marked in dark blue and the color of sampled entries remains uncharged.); (b2) The estimated SLF of source 1; (b3) The estimated SLF of source 2. }
	\label{SLF}
\end{figure}

In Figs.~\ref{PSD}-\ref{SLF}, we visualize the results of an $R=2$ and $K =64$ case. We set ${M=5,N=6}$, and assume that equi-spaced frontal and horizontal slabs are sampled---i.e., $\tX^{(1)} = \tX({\cal S}_1,:,:)$ and $\tX^{(2)} = \tX(:,{\cal S}_2,:)$ are sampled (where {${\cal S}_1=\{17,33,    49,    65,    81\},{\cal S}_2=\{15,    29,    43,    57,    71,    85\}$)}.  We set $L=4$ and assume that $R$ is known.
The ground-truth PSDs and estimated PSDs (after removing permutations and rescaling the maximal amplitude of every PSD to 1) of the 2 sources are plotted in Fig.~\ref{PSD}. The estimated SLFs are shown in Fig. \ref{SLF} (a1-b3).
One can see that both the PSDs and SLFs are reasonably recovered, under such a challenging scenario.

We also evaluate NAE$_{\C}$, NAE$_{\S}$ and NAE$_{\tX}$ versus $L$ under different $M,N,R$. The results are presented in Table~\ref{tab:Lchange} under $K=128$. One can see that the performance of the proposed approach deteriorates as $R$ increases.
This is reasonable, since many more parameters have to be estimated under a larger $R$---whereas the number of measurements remain the same.
{ One can also see that the choice of $L$ does affect the recovery accuracy to a certain extent, but not heavily. The chosen $L$ controls the approximation quality for the SLFs. The tradeoff is that a small $L$ may result in a coarse approximation for the SLFs, but a large $L$ makes computation harder.}

\begin{tiny}
\begin{table}[!htbp]
	
	\scriptsize
	\caption{NAEs under Different $L,R$.}\label{tab:Lchange}
	\begin{center}
		\begin{tabular}{ |c|c|c|c|c|c| }
			\hline
			NAE$_{\C}$&	$L=2$	&$L=3$&	$L=4$	&$L=5$\\ \hline
			$R=2,M=15,N=6$& 0.0038 &   0.0009 &   0.0023  &  0.0010\\ \hline
			$R=3,M=15,N=6$&   0.0067 &   0.0065 &   0.0038  &  0.0056\\ \hline
			$R=5,M=15,N=6$&	    0.0219  &  0.0198  &  0.0270  &  0.0214\\  \hline
			$R=8,M=15,N=6$&  0.0731   & 0.0856  &  0.0620  &  0.0923\\  \hline
			NAE$_{\S}$&	$L=2$	&$L=3$&	$L=4$	&$L=5$\\ \hline
			$R=2,M=15,N=6$&     0.2335  & 0.2317 &   0.2393 &   0.2456\\ \hline
			$R=3,M=15,N=6$&  0.2433 &   0.2515 &   0.2358 &   0.2143\\ \hline
			$R=5,M=15,N=6$&	 0.2436  &  0.2331 &   0.2528  &  0.2539\\ \hline
			$R=8,M=15,N=6$&    0.2656 &   0.2786 &   0.2611   & 0.2664\\ \hline
			NAE$_{\tX}$&	$L=2$	&$L=3$&	$L=4$	&$L=5$\\ \hline
			$R=2,M=15,N=6$&     0.2087 &   0.2114 &  0.2169  &  0.2253\\ \hline
			$R=3,M=15,N=6$&   0.2039  &  0.2154 &   0.2025 &   0.2016\\ \hline
			$R=5,M=15,N=6$& 0.2118 &   0.2084 &   0.2285  &  0.2228\\ \hline
			$R=8,M=15,N=6$& 0.2291 &   0.2442 &   0.2265  &  0.2361\\ \hline
		\end{tabular}
	\end{center}
\end{table}
\end{tiny}

Next, we compare our algorithm with some baseline methods. In \cite{romero2017learning}, the authors proposed a \textit{semi-parametric regression} (SR) to disaggregate the radio map and estimate the SLFs of each transmitters from multiple measurements per sensor.
There, the PSD of each transmitter is assumed known. The SLFs of the emitters can be estimated under this assumption using a regression based approach.
For SR, we only { use} NAE$_{\S}$ and NAE$_{\tX}$ as its performance metrics (since SR assumes that $\bm C$ is known).
Note that SR also works with quantized measurements, which we do not consider in this work.
Another baseline method is the TPS interpolation technique proposed in \cite{ureten2012comparison}.
TPS is designed for single-band cartography. Hence, we perform TPS on each of the frontal slabs of the radio map tensor, i.e., $\tX(:,:,k)$ for $k=1,\ldots,K$. We then apply the {\sf LL1} decomposition on the TPS-estimated $\widehat{\tX}$ to disaggregate the PSDs and SLFs and observe the results.
We also implement the group lasso splines (GLS) method in \cite{bazerque2011group} as another baseline.
GLS involves a basis selection step for representing the PSDs using an over-complete dictionary; we simplify their problem setting by giving GLS the real basis of the simulated PSDs, i.e., $\text{sinc}^2\left(\frac{k-f_i^r}{w_i^r}\right)$ for $f_i\in{\cal B}$, where ${\cal B}$ is an index set that includes the true locations of the occupied frequency bins by emitter $k$. This way, GLS deals with an easier task.
GLS also does not estimate the emitter-level information and we apply {\sf LL1} decomposition on $\widehat{\tX}$ as in the TPS case.

	\begin{tiny}
	\begin{table}
\scriptsize
		\caption{Performance under $L=4,R=3,N=6$.}\label{tab:Mchange}
		\begin{center}
			\begin{tabular}{ |c|c|c|c|c|c| }
				\hline
				\multirow{2}{4em}{Algorithm}&\multirow{2}{4em}{Measure} & \multicolumn{4}{|c|}{$M$} \\ \cline{3-6}
				&&5&10&15&20\\
				\hline
				\multirow{4}{4em}{Proposed}& NAE$_{\C}$ &    0.0381 &   0.0088    & 0.0055   &0.0051\\ \cline{2-6}
				&NAE$_{\S}$ &     0.4717  &  0.3420 &   0.2474  &  0.1905 \\  \cline{2-6}
				&NAE$_{\tX}$ &   0.3889 &   0.3017&    0.2194 &   0.1693 \\ \cline{2-6}
				&Running time(s)  & 1.1084 &   1.1716&    1.4226 &   1.6292\\ \cline{2-6}

				
				\hline
				\multirow{3}{4em}{SR }&NAE$_{\S}$ &   0.4580    &0.3208   & 0.2281   & 0.1884\\ \cline{2-6}
				&NAE$_{\tX}$&  0.3795  &  0.2824 &   0.2083 &   0.1760\\ \cline{2-6}
				&Running time(s)&0.2883   & 0.4581 &   0.6992   & 0.9900\\
				\hline
				\multirow{4}{4em}{TPS }&NAE$_{\C}$ &0.5995 &   0.0246&    0.0117 &   0.0110 \\ \cline{2-6}
                &NAE$_{\S}$ &     0.9922  &  0.5430&    0.3753  &  0.3249\\  \cline{2-6}
				&NAE$_{\tX}$ &  0.6671  &  0.4487&    0.3226 &   0.2492\\ \cline{2-6}
				&Running time(s)& 13.3200 &  21.3532&   31.2258&   43.0450\\
				\hline
				\multirow{3}{4em}{GLS }&NAE$_{\C}$ & 0.5741 &   0.0216&    0.0427&    0.0251 \\ \cline{2-6}
                &NAE$_{\S}$ &    0.9915 &   0.5518 &   0.4487&    0.4402\\  \cline{2-6}
				&NAE$_{\tX}$ &  0.6549 &   0.4746&    0.3196&    0.2793\\ \cline{2-6}
				&Running time(s)& 32.1612 &  59.2353&  113.6869 & 195.5264\\
				\hline
			\end{tabular}
		\end{center}
		\vspace{-.5cm}
	\end{table}
\end{tiny}

Table \ref{tab:Mchange} shows the results under $L=4,R=3,N=6,K=128$ and various { $M$s}.
In general, the performance improves for all the algorithms when $M$ increases, since more samples are available.
SR exhibits the best performance in terms of running time---since it does not need to estimate $\C$. The NAE$_{\S}$ and NAE$_{\tX}$ of the proposed algorithm  approximate those of SR.
The  { NAE$_{\S}$s}  of TPS and GLS are worse than that of the proposed coupled tensor approach, perhaps because they did not exploit the signal model information or the low-rank structure of the SLFs in \eqref{eq:sigmod_tensor}. 

\begin{figure}
	\centering
	\includegraphics[width=.9\linewidth]{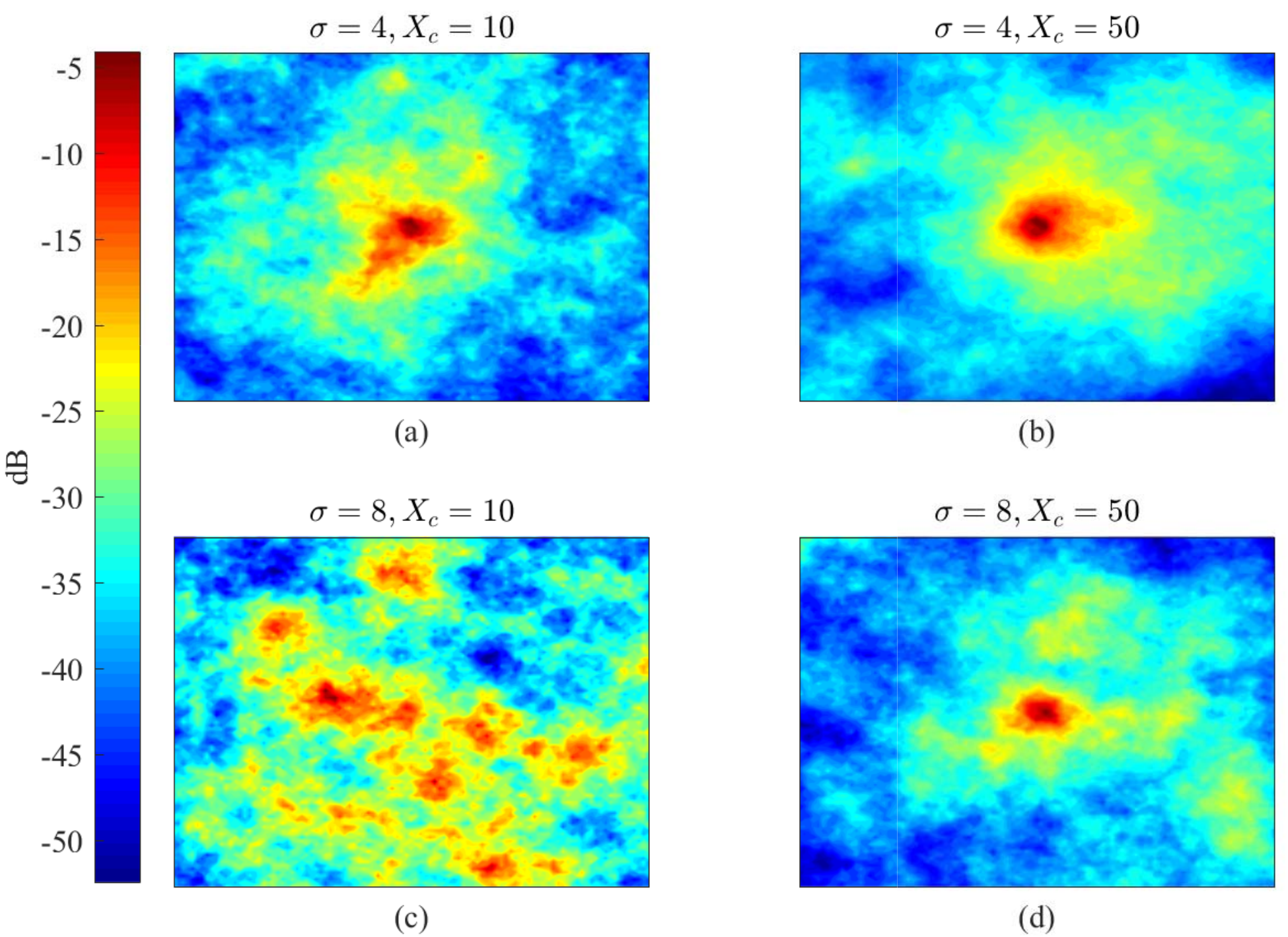}
	\caption{Shadowing effects of an SLF under various parameter settings.}
	\label{fig:scattering}
	\vspace{-.5cm}
\end{figure}

Tables~\ref{XcSigma4}-\ref{XcSigma8} show the performance of the algorithms under different shadowing  effect.
This is of interest, since if the  shadowing effect is severe, the low-rank assumption of $\S_r$ may be grossly violated.
In particular, we test the algorithms under different { $X_c$s} and { $\sigma$} in \eqref{eq:Xc}. Note that a larger $X_c$ means milder  shadowing effect while a larger { $\sigma$} corresponds to more severe shadowing; see Fig.~\ref{fig:scattering}. For outdoor environments, $X_c$ is typically between 50 and 100 \cite{goldsmith2005wireless}. Nonetheless, we test the algorithms under 10 to 100, to check if the algorithms can work under serious  shadowing. One can see in both tables, the proposed approach outperform the baselines under most cases, even if the  shadowing effect is heavy (e.g., $X_c=10,\sigma=8$)---which shows robustness against severe  shadowing.

\begin{tiny}
	\begin{table}
		\scriptsize
		\caption{Performance w.r.t different $X_c$ under $M=15,N=6,K=128,R=3,\sigma=4$.}\label{XcSigma4}
		\begin{center}
			\begin{tabular}{ |c|c|c|c|c|c|c| }
				\hline
				\multirow{2}{4em}{Algorithm}&\multirow{2}{4em}{Measure} & \multicolumn{5}{|c|}{$X_c$} \\ \cline{3-7}
				&&10&30&50&70&100\\
				\hline
				\multirow{3}{4em}{Proposed}&NAE$_{\C}$ &   0.0074&	0.0056&	0.0044&	0.0038&	0.0026
\\ \cline{2-7}
				&NAE$_{\S}$ &     0.2979&	0.2201&	0.2063&	0.1826&	0.1855\\  \cline{2-7}
				&NAE$_{\tX}$ &   0.2732&	0.1951&	0.1841&	0.1686&	0.1703 \\ \cline{2-7}
				\hline
				\multirow{2}{4em}{SR}&NAE$_{\S}$ &  0.305&	0.209&	0.1944&	0.196&	0.1667 \\ \cline{2-7}
				&NAE$_{\tX}$ &   0.2742&	0.191&	0.1761&	0.1701&	0.1465 \\ \cline{2-7}
\hline
				\multirow{3}{4em}{TPS}&NAE$_{\C}$ &   0.0129&	0.0051&	0.0116&	0.0064&	0.0074
\\ \cline{2-7}
				&NAE$_{\S}$ &     0.443&	0.3582&	0.4145&	0.3246&	0.356\\  \cline{2-7}
				&NAE$_{\tX}$ &
0.4057&	0.3035&	0.3116&	0.294&	0.2989 \\ \cline{2-7}
\hline
				\multirow{3}{4em}{GLS}&NAE$_{\C}$ &   0.0143  &  0.0092  &  0.0075&    0.0092&    0.0052
\\ \cline{2-7}
				&NAE$_{\S}$ &   0.4717  &  0.3648  &  0.3653 &   0.3875&    0.3450\\  \cline{2-7}
				&NAE$_{\tX}$ &
0.3936 &   0.3045  &  0.3251 &   0.3189 &   0.2891 \\ \cline{2-7}

\hline

			\end{tabular}
		\end{center}
		\vspace{-.5cm}
	\end{table}
\end{tiny}

\begin{tiny}
	\begin{table}
		\scriptsize
		\caption{Performance w.r.t different $X_c$ under $M=15,N=6,K=128,R=3,\sigma=8$.}\label{XcSigma8}
		\begin{center}
			\begin{tabular}{ |c|c|c|c|c|c|c| }
				\hline
				\multirow{2}{4em}{Algorithm}&\multirow{2}{4em}{Measure} & \multicolumn{5}{|c|}{$X_c$} \\ \cline{3-7}
				&&10&30&50&70&100\\
				\hline
				\multirow{3}{4em}{Proposed}&NAE$_{\C}$ &  0.0181&	0.0109&	0.0076&	0.0094&	0.009
\\ \cline{2-7}
				&NAE$_{\S}$ &     0.5328&	0.3482&	0.2768&	0.2543&	0.2354\\  \cline{2-7}
				&NAE$_{\tX}$ &   0.4736&	0.3143&	0.2544&	0.2282&	0.2132\\ \cline{2-7}
				\hline
				\multirow{3}{4em}{SR}&NAE$_{\S}$ &  0.5156&	0.332&	0.2717&	0.2543&	0.2092 \\ \cline{2-7}
				&NAE$_{\tX}$ &   0.4601&	0.3036&	0.2471&	0.2292&	0.197 \\ \cline{2-7}
\hline
				\multirow{3}{4em}{TPS}&NAE$_{\C}$ &   0.3692&	0.3633&	0.0325&	0.0202&	0.0071
\\ \cline{2-7}
				&NAE$_{\S}$ &    0.8264&	0.6841&	0.4503&	0.397&	0.385\\  \cline{2-7}
				&NAE$_{\tX}$ &
0.6117&	0.379&	0.354&	0.3399&	0.3411\\ \cline{2-7}
\hline
				\multirow{3}{4em}{GLS}&NAE$_{\C}$ &   0.0452  &  0.0407  &  0.0575  &  0.0155 &   0.0123
\\ \cline{2-7}
				&NAE$_{\S}$ &   0.6642  &  0.5023   & 0.4213   & 0.4081 &   0.3892\\  \cline{2-7}
				&NAE$_{\tX}$ &
 0.5758  &  0.4080  &  0.3331  &  0.3115   & 0.3165\\ \cline{2-7}
\hline

			\end{tabular}
		\end{center}
		\vspace{-.5cm}
	\end{table}
\end{tiny}

\subsection{Static Sensors and Fiber Sampling}
In this subsection, we consider the fiber sampling pattern of the scattered sensor deployment case.  We define the sampling ratio $\rho=\frac{\text{\# of sampled entries}}{IJK}.$ First, we give the results of { NAEs} versus different sampling ratio, under $K=128,R=3,L=4$.
In random fiber sampling, $\rho$ varies from 0.10 to 0.25 with interval 0.05.

\begin{figure}[!htbp]
	\centering
	\includegraphics[width=0.97\linewidth]{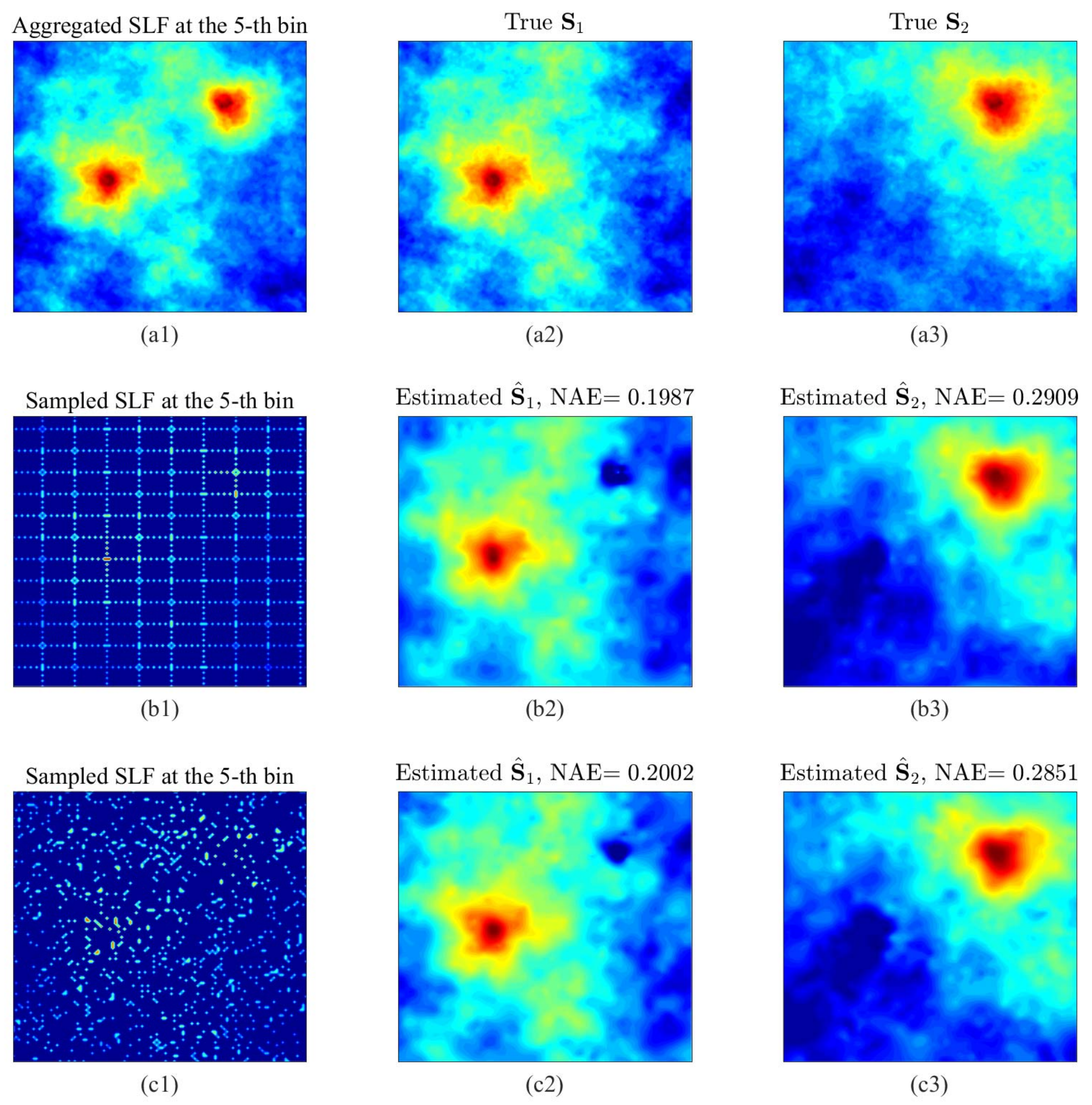}
	\caption{ Performance under regular and random sampling with {$\rho=0.10$.   (a1) The ground-truth aggregated SLF at the 5-th frequency bin; (a2) - (a3) The ground-truth SLFs. (b1) regular sampling pattern similar as that in Fig.~\ref{MultiTensorFig}. (b2)-(b3): estimated SLFs by the proposed approach. (c1) random sampling pattern. (c2)-(c3) by the proposed approach.}}
	\label{SLF2}
\end{figure}

We visualize the estimated SLFs in Fig. \ref{SLF2}. Here, $R = 2$ transmitters and $K = 64$ frequency bands are considered.
The ground-truth SLFs are the same as those in Fig.~\ref{SLF}.
The fiber sampling pattern in (a1) is similar to the one that we saw in Fig.~\ref{MultiTensorFig}. One can see that the SLFs are well recovered.
A random fiber sampling pattern is considered in Fig. \ref{SLF} (b1), where approximately 10\% of the fibers are sampled.
One can see that our algorithm also works for random sampling, which supports our claim in Proposition~\ref{FibSampTheorem}.

\begin{tiny}
\begin{table}
\scriptsize
\caption{Performance under $L=4,R=3,K=128$.}\label{tab:fiber_random}
\begin{center}
\begin{tabular}{ |c|c|c|c|c|c|}
\hline
\multirow{2}{4em}{Algorithm}&\multirow{2}{4em}{Measure} & \multicolumn{4}{|c|}{$\rho$} \\ \cline{3-6}
&&0.1&0.15&0.2&0.25\\
\hline
\multirow{4}{4em}{Proposed}&NAE$_{\C}$  & 0.0176&    0.0154 &   0.0198  &  0.0188 \\ \cline{2-6}
&NAE$_{\S}$&  0.2152  &  0.1721&    0.1583 &   0.1329  \\ \cline{2-6}
&NAE$_{\tX}$ &  0.1983  &  0.1629&    0.1400 &   0.1190
\\ \cline{2-6}
&Running time(s) & 93.5609  & 92.3041 &  93.0684 &  93.6724   \\
\hline
\multirow{3}{4em}{SR }&NAE$_{\S}$ &   0.1846  &  0.1537 &   0.1345&    0.1231\\ \cline{2-6}
&NAE$_{\tX}$ &     0.1769 &   0.1440   & 0.1267   & 0.1158\\ \cline{2-6}
&Running time(s)&  0.3684 &   0.5481  &  0.7880 &   1.1011\\
\hline
\multirow{4}{4em}{TPS }&NAE$_{\C}$ &    0.0120  &  0.0153  &  0.0091&    0.0086 \\ \cline{2-6}
&NAE$_{\S}$ &    0.4388 &   0.3688  &  0.3191  &  0.2685\\ \cline{2-6}
&NAE$_{\tX}$ &   0.4063 &   0.3301&    0.2786 &   0.2424 \\ \cline{2-6}
&Running time(s)&    14.9322 &  30.1523 &  48.6299&   54.0290\\
\hline
\multirow{4}{4em}{GLS }&NAE$_{\C}$ &  0.0306&    0.0087&    0.0063&    0.0076 \\ \cline{2-6}
&NAE$_{\S}$ &     0.3452 &   0.2558&    0.2287 &   0.1980\\  \cline{2-6}
&NAE$_{\tX}$ &   0.3061 &   0.2328&    0.2059&    0.1764\\ \cline{2-6}
&Running time(s)&  21.8072&   49.2587&  111.7062&  203.3006\\
\hline
\end{tabular}
\end{center}
	\vspace{-.5cm}
\end{table}
\end{tiny}

Table~\ref{tab:fiber_random} shows the { NAEs} performance under different sampling ratios. The fibers are sampled uniformly at random.
One can see that the estimation accuracy of the proposed approach is very promising, whose $\bm S$-estimation and $\X$-estimation  accuracy approach the performance of SR---but SR assumes that $\bm C$ is known, while the proposed approach is a completely blind method.
In addition, our method for estimating $\S$ and $\X$ is more accurate compared to TPS and GLS, which is similar to what we have observed in the slab sampling case.

\subsection{Performance under Noise}
We also consider a case where the measurements are noisy, i.e.,
$
\tilde{\tX} = \tX + \tN
$
where $\tX$ follows the model in \eqref{eq:sigmod_tensor} and the entries of $\tN$ follow the zero-mean i.i.d. Gaussian distribution.  The signal-to-noise ratio (SNR) is defined as $10\log_{10} (\Vert\tX\Vert_F^2/\Vert\tN\Vert_F^2)$. { We would like to remark that the Gaussian noise is added to test robustness of the algorithms under model mismatch. In practice, the noise may not be Gaussian since the data is nonnegative.}

Table \ref{tab:SNR_Slab} and Table \ref{tab:SNR_Fiber} show the performance of the algorithms against various SNRs, under the slab sampling and fiber sampling schemes, respectively.
The situations are similar as before: The proposed approach works fairly well in terms of estimating $\bm S_r$, $\bm c$ and $\tX$, under all SNRs. This may be because the proposed approach admits identifiability supports and explicitly utilizes the underlying signal model---which is often essential for combating noise.

\begin{tiny}
\begin{table}
\scriptsize
\caption{Performance of slab sampling under $L=4,R=3,K=128,M=15,N=6$ versus different SNR.}\label{tab:SNR_Slab}
\begin{center}
\begin{tabular}{ |c|c|c|c|c|c|}
\hline
\multirow{2}{4em}{Algorithm}&\multirow{2}{4em}{Measure} & \multicolumn{4}{|c|}{SNR(dB)} \\ \cline{3-6}
&&0&10&20&30\\
\hline
\multirow{3}{4em}{Proposed}&NAE$_{\C}$  &    0.1532 &   0.0543  &  0.0193 &   0.0098 \\ \cline{2-6}
&NAE$_{\S}$&   0.5441&    0.3494&    0.2797  &  0.2595  \\ \cline{2-6}
&NAE$_{\tX}$ &  0.6034&    0.3431 &   0.2557 &   0.2282
\\
\hline
\multirow{2}{4em}{SR }&NAE$_{\S}$ &     0.6742 &   0.3912&    0.2711  &  0.2450\\ \cline{2-6}
&NAE$_{\tX}$ &      0.9681 &   0.4288 &   0.2654  &  0.2204\\
\hline
\multirow{3}{4em}{TPS }&NAE$_{\C}$ &      0.1993 &   0.0641   & 0.0259  &  0.0236 \\ \cline{2-6}
&NAE$_{\S}$ &     0.8855 &   0.5371  &  0.4450 &   0.4319\\ \cline{2-6}
&NAE$_{\tX}$ &    10.2674  &  3.4584&    1.2773 &   0.6349 \\
\hline
\multirow{3}{4em}{GLS }&NAE$_{\C}$ &   0.0271   & 0.0119 &   0.0126  &  0.0137 \\ \cline{2-6}
&NAE$_{\S}$ &    0.8381  &  0.4849  &  0.4176 &   0.3450\\ \cline{2-6}
&NAE$_{\tX}$ &   1.0426 &   0.5175  &  0.3909  &  0.3109 \\
\hline
\end{tabular}
\end{center}
	\vspace{-.5cm}
\end{table}
\end{tiny}

\begin{tiny}
\begin{table}
\scriptsize
\caption{Performance of fiber sampling under $L=4,R=3,K=128,p=0.15$ versus different SNR.}\label{tab:SNR_Fiber}
\begin{center}
\begin{tabular}{ |c|c|c|c|c|c|}
\hline
\multirow{2}{4em}{Algorithm}&\multirow{2}{4em}{Measure} & \multicolumn{4}{|c|}{SNR(dB)} \\ \cline{3-6}
&&0&10&20&30\\
\hline
\multirow{3}{4em}{Proposed}&NAE$_{\C}$  & 0.1779 &   0.0753&    0.0232&    0.0170 \\ \cline{2-6}
&NAE$_{\S}$&   0.7947 &   0.4178 &   0.2538  &  0.2097\\ \cline{2-6}
&NAE$_{\tX}$ &   1.1255   & 0.4803  &  0.2500&    0.2007\\
\hline
\multirow{2}{4em}{SR }&NAE$_{\S}$ &       0.7305  &  0.3856  &  0.2590    &0.2128\\ \cline{2-6}
&NAE$_{\tX}$ &       1.1170  &  0.4540  &  0.2619  &  0.2053\\
\hline
\multirow{3}{4em}{TPS }&NAE$_{\C}$ &   0.1911  &  0.0820 &   0.0343  &  0.0161 \\ \cline{2-6}
&NAE$_{\S}$ &    0.9055   & 0.4974 &   0.3382 &   0.3009\\ \cline{2-6}
&NAE$_{\tX}$ &  13.9170  &  4.3835&    1.6014 &   0.6653\\
\hline
\multirow{3}{4em}{GLS }&NAE$_{\C}$ &  0.0211  &  0.0111 &   0.0068  &  0.0077 \\ \cline{2-6}
&NAE$_{\S}$ &  0.8396   & 0.4830  &  0.3312   & 0.2885\\ \cline{2-6}
&NAE$_{\tX}$ &    1.1745   & 0.5140   & 0.3259 &   0.2564\\
\hline
\end{tabular}
\end{center}
	\vspace{-.5cm}
\end{table}
\end{tiny}

\subsection{Real-data Experiment}

\begin{figure}[!hbt]
	\centering
	\includegraphics[width=1\linewidth]{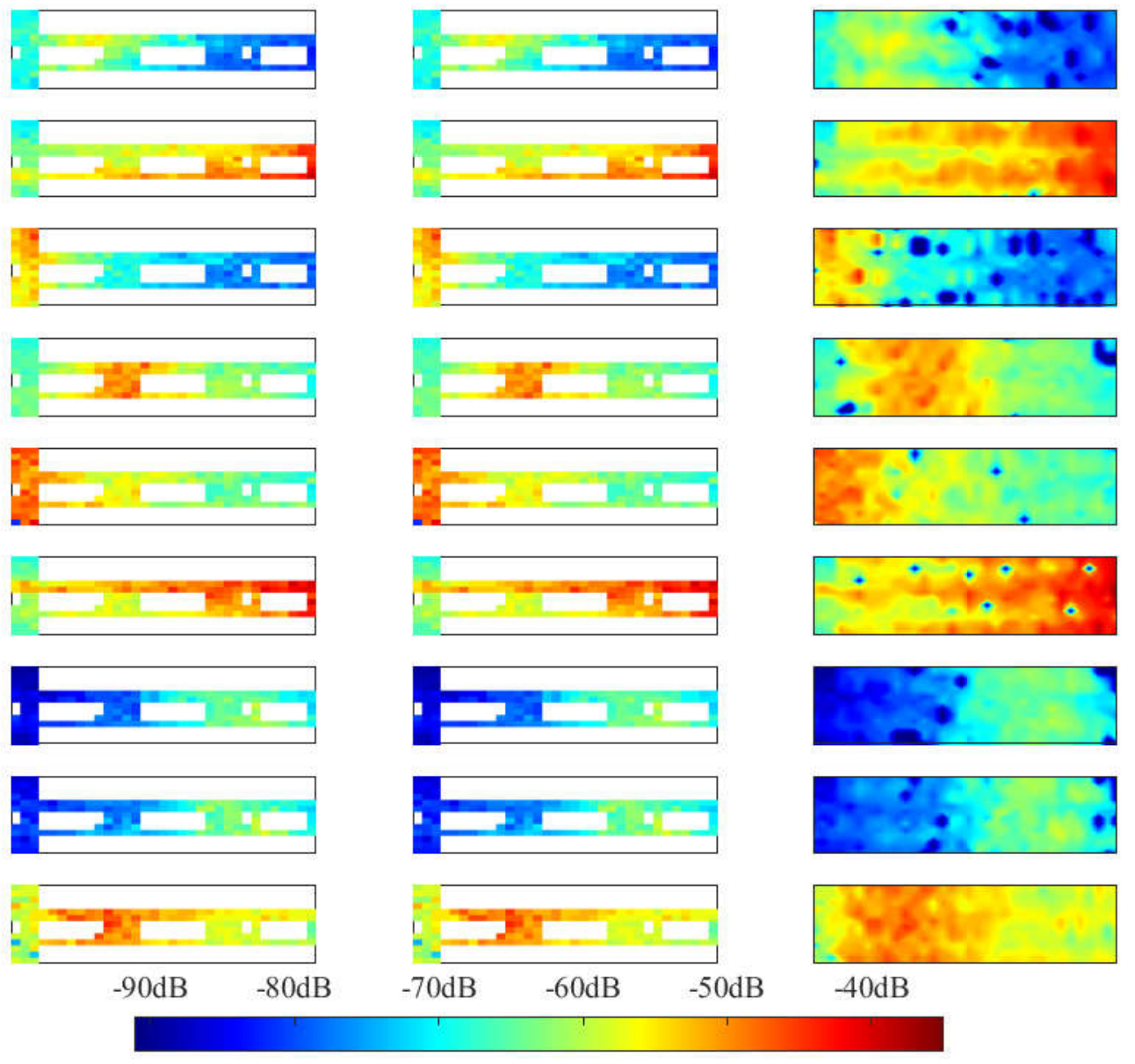}\\
	\caption{{ Recovered $\tX(:,:,k)$ using real data ($R=9,L=4$) by the proposed method. (Left) Original data; (center) Estimated data on observed locations;
			(right) Estimated data of all locations.}}
	\label{fig:RealData_BTD_R9_L4}
\end{figure}

In this section, we test our algorithm on the real data collected in \cite{mannheim-compass-20080411}.
There, the measurements are taken in a $14\times 34$ m$^2$ indoor area over 9 frequencies spanning from 2.412GHz to 2.462GHz.
The complete RF tensor is heavily downsampled [see Fig.~\ref{fig:RealData_BTD_R9_L4} (left)].
To be specific,
the complete RF tensor has a size of $14\times 34\times 9$. Three horizontal slabs are observed. Three vertical slabs are observed except for $\tX(6,1,:)$ and $\tX(7,1,:)$.	We apply our slab sampling-based methods (where $\tX(6,1,:)$ and $\tX(7,1,:)$ are assigned with the mean of their adjacent elements). Two subtensors $\tX^{(1)}=\tX(:,1:3,:)$ and $\tX^{(2)}=\tX([4,8,9],:,:)$ are thus formed. Note that we do not have the ground truth of the complete tensor. We also do not know the number of emitters in the region. Hence, we follow \cite{bazerque2011group} to offer qualitative evaluations; i.e., we plot the recovered RF tensor within the observed region and unobserved region to visually compare with the original data. To run our algorithm, we set $R=9$ and $L=4$.

Our recovered $\tX(:,:,k)$ seem to be visually smoother relative to the method in \cite{bazerque2011group} (see Fig.~8 in \cite{bazerque2011group}). This might be because the {\sf LL1} model explicitly enforces spatial smoothness (via enforcing $\S_r$ to have low rank).

\color{black}


\section{Conclusion}
In this work we proposed a novel coupled block-term tensor decomposition framework to tackle the radio map disaggregation problem. Unlike most of the existing cartography methods that are heuristic-driven, the proposed framework admits recoverability guarantees of each emitter's radio map.
In addition, the framework provably works under a number of systematic and random sampling schemes, and thus allows system designers to handle situations where sensor deployment is subject to various restrictions or regulations.
Extensive simulations under heavy-shadowing scenarios show that the proposed method is promising for enhancing RF awareness of the sensing systems.

{ In terms of future work, one potential direction is to study the frequency-selective fading scenario, under which the {\sf LL1} model does not hold. This may require more complex tensor models, e.g., those in \cite{chatzichristos2019blind,guo2012uni,stegeman2010uniqueness}. In addition, another direction may be large-scale, online coupled {\sf LL1} decomposition algorithm design incorporating prior information---which was shown effective in other tensor models in noisy situations \cite{joneidi2019primary}. Last but not least, one may take temporal variations into consideration, to accommodate fast changing scenarios.}

\appendices

%

\section{Proof of Theorem \ref{CoupledBTD}}\label{app_t3}
In the noiseless case where  (\ref{eq:Xslab}) and (\ref{Xsub}) hold, we note that the optimal solutions to Problem~\eqref{eq:coupledBTD} should make the two terms zero, when the noise is absent. Let $(\widehat{\A},\widehat{\B},\widehat{\C})$ denote any optimal solution of Problem~\eqref{eq:coupledBTD} and $(\A,\B,\C)$ denote the ground-truth. We also define $\S_r=\A_r\B_r^\top$ and $\widehat{\S}_r=\widehat{\A}_r\widehat{\B}_r^\top$. We aim to prove that $\{\widehat{\S}_r,\widehat{\vc}_r\}_{r=1}^R$ is essentially the ground-truth  $\{\S_r,\vc_r\}_{r=1}^R$ up to trivial ambiguities.

We show the case where condition (1) holds---the case under condition (2) shares exactly the same proof by the role symmetry of $M$ and $N$.  Note that $\tX^{(1)}\in\bbR^{M\times J\times K_1}$ and that $\MP\A,\B,\C(\cS_3,:)$ are latent factors of the decomposition of $\tX^{(1)}$ in multilinear rank-$(L,L,1)$ terms. Since
$R\leq \vert\cS_3\cap\cS_4\vert\leq K_1, R+2\leq \min(\lfloor M/L \rfloor, R) + \min(\lfloor J/L\rfloor ,R ),$
by Theorem 1, $\MP\A_r\B_r^\top$ and $\vc_r(\cS_3)$ for $r=1\ldots,R$ can be identified from $\tX^{(1)}$ up to scaling and permutation ambiguities. Therefore, $\C(\cS_3,:)$ can be identified up to scaling and permutation ambiguities, i.e, $\widehat{\C}(\cS_3,:)=\C(\cS_3,:)\bPi\bLambda.$

Accordingly, we have $\widehat{\vc}_r(\cS_3)=\lambda_r\vc_{\pi_r}(\cS_3)$, where $\bLambda=\tDiag(\lambda_1,\cdots,\lambda_R)$ and $\pi_r\in\{1,\cdots,R\}$ satisfies $\bPi(\pi_r,r)=1$ for $r=1,\cdots,R$.
Note that
\begin{align}
\tX^{(1)}_3 &= [\tvec(\MP\A_1\B_1^\top),\cdots,\tvec(\MP\A_R\B_R^\top)]\C(\cS_3,:)^\top \nonumber\\
&=[\tvec(\MP\widehat{\A}_1\widehat{\B}_1^\top),\cdots,\tvec(\MP\widehat{\A}_R\widehat{\B}_R^\top)]\widehat{\C}(\cS_3,:)^\top. \label{X13}
\end{align}
Plugging $\widehat{\C}(\cS_3,:)=\C(\cS_3,:)\bPi\bLambda$ into (\ref{X13}), we have $\MP\widehat{\A}_r\widehat{\B}_r^\top = \lambda_r^{-1}\MP\A_{\pi_r}\B_{\pi_r}^\top.$ Here $\A_r,\widehat{\A}_r,\B_r,\widehat{\B}_r$ are all full column-rank matrices almost surely (since they are drawn from certain joint continuous distributions), and the row selection matrix $\MP$ is a full row-rank matrix. Hence, $\MP\A_r\in\bbR^{M\times L}$ is a submatrix of $\A_r$ and $M\ge 2L$, which means that $\MP\A_r$ is a full column-rank matrix and so is $\MP\widehat{\A}_r$. Therefore, there exists a full rank matrix $\F_r$ satisfying $\widehat{\B}_r=\B_{\pi_r}\F_r$.

Note that $\cS = \cS_3\cap\cS_4$ and $\widehat{\C}(\cS_3,:)=\C(\cS_3,:)\bPi\bLambda$. Hence, we have $\widehat{\C}(\cS,:)=\C(\cS,:)\bPi\bLambda.$ Consider the subtensor $\tX^{(3)}=\tX(:,\cS_2,\cS)$, where we have
\begin{equation}
\tX^{(3)}_3=\M_Q \C(\cS,:)^\top=\widehat{\M}_Q\widehat{\C}(\cS,:)^\top\label{X33},
\end{equation}
where
\begin{align*}
\M_Q&=[\tvec(\A_1\B_1^\top\Q^\top),\cdots,\tvec(\A_R\B_R^\top\Q^\top)]\\
\widehat{\M}_Q&=[\tvec(\widehat{\A}_1\widehat{\B}_1^\top\Q^\top),\cdots,\tvec(\widehat{\A}_R\widehat{\B}_R^\top\Q^\top)].
\end{align*}
Plugging $\widehat{\C}(\cS,:)=\C(\cS,:)\bPi\bLambda$ into (\ref{X33}), we have $\widehat{\A}_r\widehat{\B}_r^\top\Q^\top = \lambda_r^{-1}\A_{\pi_r}\B_{\pi_r}^\top\Q^\top$ since $\C(\cS,:)$ is full column rank. Here $\Q$ is a row-selection matrix corresponding to ${\cal S}$. Plug $\widehat{\B}_r=\B_{\pi_r}\F_r$ into this equation, we have $\widehat{\A}_r\F_r^\top\B_{\pi_r}^\top\Q^\top = \lambda_r^{-1}\A_{\pi_r}\B_{\pi_r}^\top\Q^\top.$ Since $\B_{\pi_r}^\top\Q^\top$ is a full row rank matrix,  we can infer $\widehat{\A}_r=\lambda_r^{-1}\A_{\pi_r}\F_r^{-\top}$. Therefore,
$
\widehat{\S}_r=\widehat{\A}_r\widehat{\B}_r^\top=\lambda_r^{-1}\A_{\pi_r}\F_r^{-\top}(\B_{\pi_r}\F_r)^\top=\lambda_r^{-1}\A_{\pi_r}\B_{\pi_r}^\top=\lambda_r^{-1}\S_{\pi_r};
$
that is, we have proven that $\{\widehat{\S}_r\}_{r=1}^R$ is a scaling and permutation version of ground-truth  $\{\S_r\}_{r=1}^R$.

Next, we aim to prove that $\widehat{\C}=\C\bPi\bLambda$. In particular, if $\cS_3=\cS_4=\cS=[K]$, $\widehat{\C}=\C\bPi\bLambda$ holds since the {\sf LL1} decomposition of $\tX^{(1)}$ is essentially unique.

If ${\cal S}_3,{\cal S}_4\neq [K]$ while ${\cal S}_3\cup{\cal S}_4 =[K]$, noting that
$
\widehat{\A}_r\widehat{\B}_r^\top\Q^\top=\widehat{\S}_r\Q^\top=\lambda_r^{-1}\S_{\pi_r}\Q^\top,
$
we have
\begin{align*}
\widehat{\M}_Q&=[\tvec(\widehat{\A}_1\widehat{\B}_1^\top\Q^\top),\cdots,\tvec(\widehat{\A}_R\widehat{\B}_R^\top\Q^\top)\\
& = [\lambda_1^{-1}\tvec(\A_{\pi_1}\B_{\pi_1}^\top\Q^\top),\cdots,\lambda_R^{-1}\tvec(\A_{\pi_R}\B_{\pi_R}^\top\Q^\top)]\\
&=\M_Q\bPi\bLambda^{-1}.
\end{align*}
Consider the subtensor $\tX^{(2)}$, we have
\begin{equation}
\tX^{(2)}_3=\M_Q \C(\cS_4,:)^\top=\widehat{\M}_Q\widehat{\C}(\cS_4,:)^\top.\label{X23}
\end{equation}
Plugging $\widehat{\M}_Q=\M_Q\bPi\bLambda^{-1}$ into (\ref{X23}), we obtain $\widehat{\C}(\cS_4,:)=\C(\cS_4,:)\bPi\bLambda.$ Combining with the fact that $\widehat{\C}(\cS_3,:)=\C(\cS_3,:)\bPi\bLambda$ and $\cS_3\cup\cS_4=[K]$, we have $\widehat{\C}=\C\bPi\bLambda$. 

\section{Proof of Theorem \ref{MultiTensor}} \label{app_t4}
In the noiseless case, we note that the optimal solutions to Problem (\ref{eq:coupledBTD_fiber}) should make the two terms zero. Let $(\widehat{\A},\widehat{\B},\widehat{\C})$ denote any optimal solution of Problem  (\ref{eq:coupledBTD_fiber}) and $(\A,\B,\C)$ denote the ground-truth. Note that
\begin{align*}
\tX\dd&=\sum_{r=1}^R(\A_r(\CI\dd,:)\B_r(\CJ\dd,:)^\top)\circ \C(\CK\dd,:)\\
&=\sum_{r=1}^R(\widehat{\A}_r(\CI\dd,:)\widehat{\B}_r(\CJ\dd,:)^\top)\circ \widehat{\C}(\CK\dd,:).
\end{align*}
By Theorem \ref{BTD_theorem_IJKR}, one can see that every $\tX\dd$ admits a unique BTD, which means that $\widehat{\C}(\CK\dd,:)$ is a column permutated   and scaled version of  $\C(\CK\dd,:)$. Therefore, there exists  a  permutation matrix $\bPi\dd\in\bbR^{R\times R}$ and a  nonsingular
diagonal matrix $\bLambda\dd\in\bbR^{R\times R}$ such that
\[
\widehat{\C}(\CK\dd,:)=\C(\CK\dd,:)\bPi\dd\bLambda\dd.
\]
For any two coupled tensors $\tX\id$ and $\tX\idp$, we have
\begin{align*}
\widehat{\C}(\CK\id,:)&=\C(\CK\id,:)\bPi\id\bLambda\id,\\
\widehat{\C}(\CK\idp,:)&=\C(\CK\idp,:)\bPi\idp\bLambda\idp.
\end{align*}
We aim to prove that $\bPi\id=\bPi\idp$ and $\bLambda\id=\bLambda\idp$ hold for $d=1,\cdots,D-1$. Define $\CR\id\doteq\CK\id\cap\CK\idp$. Considering the common part $\widehat{\C}(\CR\id,:)$, we have
\begin{small}
	\begin{align*}
	\widehat{\C}(\CR\id,:)&=\widehat{\C}(\CK\id\cap\CR\id,:)=\C(\CR\id,:)\bPi\id\bLambda\id,\\
	\widehat{\C}(\CR\id,:)&=\widehat{\C}(\CK\idp\cap\CR\id,:)=\C(\CR\id,:)\bPi\idp\bLambda\idp.
	\end{align*}
\end{small}
Define $\D\id=\bPi\id\bLambda\id$ and $\vd_r\id=\D\id(:,r)$. We have
\begin{equation}\label{CD}
\C(\CR\id,:)\D\id=\C(\CR\id,:)\D\idp.
\end{equation}
Consider the $r$-th column of (\ref{CD}). We have
\[
\C(\CR\id,:)(\vd_r\id-\vd_r\idp)=\zero.
\]
Note that $\vd_r\id\in\bbR^R$ has only one non-zero element, then $\vd_r\id-\vd_r\idp$ has at most two non-zeros elements. Note that $\vert\CR\id\vert\ge 2$, which means that any two columns of $\C(\CR\id,:)$ are not proportional almost surely. Therefore, $\vd_r\id=\vd_r\idp$ holds for all $r=1,\cdots,R$, and thus
\[
\bPi\id\bLambda\id=\bPi\idp\bLambda\idp
\]
holds. Therefore,
\[
\bPi\id = \bPi\idp, \bLambda\id=\bLambda\idp
\]
for all $d=1,\cdots,D-1$. Denote $\bPi=\bPi\io$ and $\bLambda=\bLambda\io$. Concatenating all $\widehat{\C}(\CK\id,:)=\C(\CK\dd,:)\bPi\bLambda$ for $d=1,\cdots,D$,  we have
\[
\widehat{\C}=\widehat{\C}(\CK\io\cup\cdots\cup\CK^{(i_D)}:)=\C\bPi\bLambda.
\]

Next, we aim to prove that $\widehat{\S}=\S\bPi\bLambda^{-1}$ where
\begin{align*}
\S&=[\tvec(\S_1),\cdots,\tvec(\S_R)],\\
\widehat{\S}&=[\tvec(\widehat{\S}_1),\cdots,\tvec(\widehat{\S}_R)].
\end{align*}
and $\S_r=\A_r\B_r^\T$ and $\widehat{\S}_r=\widehat{\A}_r\widehat{\B}_r^\T$.
Note that
\begin{align*}
\tX_3\dd & =\underbrace{[\cdots,\tvec(\S_r(\CI\dd,\CJ\dd)),\cdots]}_{\S\dd}\C(\CK\dd,:)^\T\\
&=\underbrace{[\cdots,\tvec(\widehat{\S}_r(\CI\dd,\CJ\dd)),\cdots]}_{\widehat{\S}\dd}\widehat{\C}(\CK\dd,:)^\T,\\
\end{align*}
and $\widehat{\C}(\CK\dd,:)=\C(\CK\dd,:)\bPi\bLambda$. Therefore,  we have
\begin{equation}\label{ShatS}
\widehat{\S}\dd = \S\dd\bPi\bLambda^{-1},
\end{equation}
where $\bLambda=\tDiag(\lambda_1,\cdots,\lambda_R)$ and there exists $\pi_r\in\{1,\cdots,R\}$ such that $\bPi(\pi_r,r)=1$. Reshaping the $r$-th column of (\ref{ShatS}) to matrix, we have
\begin{equation}\label{Shatr}
\widehat{\S}_r(\CI\dd,\CJ\dd)=\lambda_r^{-1}\S_{\pi_r}(\CI\dd,\CJ\dd).
\end{equation}
Plugging $\S_r=\A_r\B_r^\T$ and $\widehat{\S}_r=\widehat{\A}_r\widehat{\B}_r^\T$ into (\ref{Shatr}), we have
\[
\widehat{\A}_r(\CI\dd,:)\widehat{\B}_r(\CJ\dd,:)^\T=\lambda_r^{-1}\A_{\pi_r}(\CI\dd,:)\B_{\pi_r}(\CJ\dd,:)^\T.
\]
Since both $\widehat{\A}_r(\CI\dd,:)$ and $\widehat{\B}_r(\CJ\dd,:)$ are with full column rank almost surely for all $r$ due to $\vert\CI\dd\vert\ge L$ and $\vert\CJ\dd\vert\ge L$, there exists a nonsingular matrix $\F_r\dd$ such that
\begin{align*}
\widehat{\A}_r(\CI\dd,:)& = \A_{\pi_r}(\CI\dd,:)\F_r\dd,\\
\widehat{\B}_r(\CJ\dd,:)&=\lambda_r^{-1}\B_{\pi_r}(\CJ\dd,:)(\F_r\dd)^{-\T}.
\end{align*}

Define $\CP\id=\CI\id\cap\CI\idp$ and $\CQ\id=\CJ\id\cap\CJ\idp$. Note that $\max(\vert \CP\id\vert,\vert \CQ\id\vert)\ge L$. If we assume  $\vert \CP\id\vert\ge L$ holds for $d\in\{1,\cdots,D-1\}$. Consider the common row indices of the following:
\begin{align*}
\widehat{\A}_r(\CI\id,:)& = \A_{\pi_r}(\CI\id,:)\F_r\id,\\
\widehat{\A}_r(\CI\idp,:)& = \A_{\pi_r}(\CI\idp,:)\F_r\idp.
\end{align*}
We have
\begin{align*}
\widehat{\A}_r(\CP\id,:)& = \widehat{\A}_r(\CI\id\cap\CP\id,:)=\A_{\pi_r}(\CP\id,:)\F_r\id,\\
\widehat{\A}_r(\CP\id,:)&= \widehat{\A}_r(\CI\idp\cap\CP\id,:) = \A_{\pi_r}(\CP\id,:)\F_r\idp.
\end{align*}
Then, it follows that
\[
\A_{\pi_r}(\CP\id,:)(\F_r\id-\F_r\idp)=\zero.
\]
Since $\vert\CP\id\vert\ge L$, $\A_{\pi_r}(\CP\id,:)$ is a full rank matrix and we can conclude that $\F_r\id=\F_r\idp$ for $d\in\{1,\cdots,D-1\}$. Similarly, if we assume that $\vert \CQ\id\vert\ge L$ holds for $d\in\{1,\cdots,D-1\}$, we can also conclude that $\F_r\id=\F_r\idp$ for $d\in\{1,\cdots,D-1\}$. Therefore, $\F_r\id=\F_r\idp$ holds for  $d=1,\cdots,D-1$.

Denote $\F_r=\F_r\io$. Concatenating all $\widehat{\A}_r(\CI\id,:) = \A_{\pi_r}(\CI\id,:)\F_r$, we have
\[
\widehat{\A}_r=\widehat{\A}_r(\CI\io\cup\cdots\cup\CI^{(i_D)},:)=\A_{\pi_r}\F_r.
\]
Similarly, concatenating all $\widehat{\B}_r(\CJ\dd,:)=\lambda_r^{-1}\B_{\pi_r}(\CJ\dd,:)\F_r^{-\T}$, we also have
\[
\widehat{\B}_r=\lambda_r^{-1}\B_{\pi_r}\F_{r}^{-\T}
\]
Therefore, we have
\[
\widehat{\S}_r=\widehat{\A}_r\widehat{\B}_r^\T=\lambda_r^{-1}\A_{\pi_r}\B_{\pi_r}^\T,
\]
and
\begin{align*}
\widehat{\S}&=[\tvec(\widehat{\S}_1),\cdots,\tvec(\widehat{\S}_R)]\\
&=[\lambda_1^{-1}\tvec(\S_{\pi_1}),\cdots,\lambda_R^{-1}\tvec(\S_{\pi_R})]\\
&=\S\bPi\bLambda^{-1}.
\end{align*}

\section{Proof of Theorem \ref{MultiTensor-C}}\label{app_t5}
This proof is similar to that of Theorem \ref{MultiTensor}. First, we note that the optimal solutions to Problem (\ref{eq:coupledBTD_fiber}) should make the two terms zero. Let $(\widehat{\A},\widehat{\B},\widehat{\C})$ denote any optimal solution of Problem  (\ref{eq:coupledBTD_fiber}) and $(\A,\B,\C)$ denote the ground-truth. Note that
\begin{align*}
\tX\dz&=\sum_{r=1}^R(\A_r(\CI\dz,:)\B_r(\CJ\dz,:)^\top)\circ \C(\CK\dz,:)\\
&=\sum_{r=1}^R(\widehat{\A}_r(\CI\dz,:)\widehat{\B}_r(\CJ\dz,:)^\top)\circ \widehat{\C}(\CK\dz,:).
\end{align*}
Note that $\CK\dz=[K],K\ge R$ and
\[
\min\left(\bigg\lfloor\frac{\vert\CI\dz\vert}{L}\bigg\rfloor,R\right) + \min\left(\bigg\lfloor\frac{\vert\CJ\dz\vert}{L}\bigg\rfloor,R\right)\ge R+2.
\]
By Theorem \ref{BTD_theorem_IJKR}, there exists  a  permutation matrix $\bPi\in\bbR^{R\times R}$ and a  nonsingular
diagonal matrix $\bLambda\in\bbR^{R\times R}$ such that
\[
\widehat{\C}=\C\bPi\bLambda.
\]

The proof of  $\widehat{\S}=\S\bPi\bLambda^{-1}$ is totally same as that of Theorem \ref{MultiTensor}.

\bibliographystyle{IEEEtran}
\bibliography{TSPCartography}

\clearpage

{\bf Supplementary { Material} of ``Spectrum Cartography via Coupled Block-term Tensor Decomposition''}
~\\

G. Zhang, X. Fu, J. Wang, X.-L. Zhao, and M. Hong
\section{Proof of Proposition \ref{FibSampTheorem}}\label{app_t6}
The objective function (\ref{opt_rndfib}) can be rewritten as:
\begin{equation}
\sum_{k=1}^K\Vert\tW(:,:,k)\ast\big(\tX(:,:,k)-\A{ \tilde{\C}_k}\B^\top\big)\Vert_F^2,
\end{equation}
{ where $\tilde{\C}_k=\text{BlockDiag}(c_{k,1}\one_L,\cdots,c_{k,R}\one_L)$ and $\text{BlockDiag}(\cdot)$ denotes the block diagonal matrix with the augment matrices on its diagonal blocks.}
Since $I\ge LR,J\ge LR$, the low-rank matrix factorization $\A{ \tilde{\C}_k}\B^\top$ can be written as $\A{ \tilde{\C}_k}\B^\top=\U_k\V_k^\top$, $\U_k\in\bbR^{I\times LR},\V_k\in\bbR^{J\times LR}$. Then, the optimization problem (\ref{opt_rndfib}) can be rewritten as:
\begin{align*}
\min_{\A,\B,\C} \; &\sum_{k=1}^K\Vert\tW(:,:,k)\ast\big(\tX(:,:,k)-\U_k\V_k^\top\big)\Vert_F^2\\
\text{subject to}\;&\U_k\V_k^\top=\A{ \tilde{\C}_k}\B^\top, k=1,\cdots,K.
\end{align*}
In the noiseless case,
the optimal solution should make $\Vert\tW(:,:,k)\ast\big(\tX(:,:,k)-\U_k\V_k^\top\big)\Vert_F^2$ to be zero for $k=1,\cdots,K$. This is equivalent to solving $K$ matrix completion problems, each for one slab $\tX(:,:,k)$.


In \cite{pimentel2016characterization}, the authors derived sufficient conditions  for rank-$r$ matrix completion, which is stated as follows:
\begin{Lemma}
	\cite{pimentel2016characterization} Consider a low-rank matrix $\X\in\bbR^{m\times n}$ with rank $r$, where $m\leq n$ and $r\leq m/6$. Let $0<\epsilon\le 1$ be given. Each column of $\X$ is sampled uniformly at random, with at least $l$ entries observed. Also assume that $l$ satisfies
	\[
	l \ge \max\bigg\{12\log\bigg(\frac{d}{\epsilon}+1\bigg),2r\bigg\}.
	\]
	Then with probability at least $1-\epsilon$, the incomplete matrix is uniquely completable if $n\ge (r+1)(m-r)$.
\end{Lemma}
Note that $\tX(:,:,k)$ is a rank-$LR$ matrix almost surely if $\{\A_r,\B_r\}_r$ and $\C$ are drawn from any joint continuous distribution and noise is absent.
Applying the above rank-$r$ matrix completion theorem to rank-$LR$ matrix $\tX(:,:,k)$ completion, one can conclude that $\tX(:,:,k)$ can be be recovered under the conditions specified in Fact \ref{FibSampTheorem}. Once $\tX$ is recovered, the remaining problem is a rank-{ $(L,L,1)$} BTD factorization problem. Using Theorem 1, we can reach the conclusion of Fact 1.

\section{Update $\A,\B,\C$ for Solving ~\eqref{eq:coupledBTD}}\label{app_BC}
\subsection{Solution of ~\eqref{Aeq:BCD}}
{ The optimality condition of the subproblem w.r.t. $\A$ when we use BCD to solve Problem~\eqref{eq:coupledBTD} is a system of linear equations. To see this, let us first simplify the notations and} rename different parts in the { optimality condition of}~\eqref{Aeq:BCD} as follows:
\begin{align*}\label{eq:Aequal}
&\underbrace{\MP^\top\MP}_{\H_1}\A\underbrace{(\R_1\C\odot_p\B)^\top(\R_1\C\odot_p\B)}_{\H_2}+ \nonumber\\
&\quad\quad \underbrace{\I_I}_{\H_3}\A\underbrace{\left[(\R_2\C\odot_p\Q\B)^\top(\R_2\C\odot_p\Q\B) + \lambda_1\I_{LR}\right]}_{\H_4}  + \nonumber\\
&=  \underbrace{\MP^\top\X^{(1)\top}_1(\R_1\C\odot_p\B) + \X^{(2)\top}_1(\R_2\C\odot_p\Q\B)}_{\H_5}.
\end{align*}
This simplifies the above equation as
$$\H_1\A\H_2+\H_3\A\H_4=\H_5.$$
To find $\A$ from the Sylvester equation $\H_1\A\H_2+\H_3\A\H_4=\H_5$, we propose to employ the extended Bartels-Stewart method \cite{gardiner1992solution}.   

\subsection{Update $\B,\C$ for Solving ~\eqref{eq:coupledBTD}}
The factor $\B$ can be updated via solving the following equation:
\begin{align*}
&\Q^\top\Q\B[(\R_2\C\odot_p\A)^\top(\R_2\C\odot_p\A)]\\
+& \B\left[(\R_1\C\odot_p\MP\A)^\top(\R_1\C\odot_p\MP\A) + \lambda_2\I_{LR}\right]\nonumber\\
=&[\X^{(1)\top}_2(\R_1\C\odot_p\MP\A) + \Q^\top\X^{(2)\top}_2(\R_2\C\odot_p\A)].
\end{align*}
Similarly, $\C$ can be updated via solving the following equation:
\begin{align*}
&\R_1^\top\R_1\C\M_1^\top\M_1 + \R_2^\top\R_2\C\M_2^\top\M_2+\lambda_3\C \nonumber\\
=&\R_1^\top\X^{(1)\top}_3\M_1 + \R_2^\top\X^{(2)\top}_3\M_2,
\end{align*}
where
\begin{align*}
\M_1&=[(\B_1\odot\MP\A_1)\one_{L},\cdots,(\B_R\odot\MP\A_R)\one_{L}]\in\bbR^{I_1J\times R},\\
\M_2&=[(\Q\B_1\odot\A_1)\one_{L},\cdots,(\Q\B_R\odot\A_R)\one_{L}]\in\bbR^{IJ_1\times R}.
\end{align*}

\section{Update $\A,\B,\C$ for Solving ~\eqref{opt_rndfib}}\label{app_ABCfiber}
The $i$th row of $\A$ can be updated by:
\begin{align*}
\A(i,:) &= \Y_1(:,i)^\top(\C\odot_p\B)\\
&\left[
(\C\odot_p\B)^\top\Diag(\W_1(:,i))(\C\odot_p\B) + \lambda_1 \I_{LR}\right]^{-1}.
\end{align*}
The $j$th row of $\B$ can be updated by:
\begin{align*}
\B(j,:) &= \Y_2(:,j)^\top(\C\odot_p\A)\\
&\left[
(\C\odot_p\A)^\top\Diag(\W_2(:,j))(\C\odot_p\A)+ \lambda_2 \I_{LR}\right]^{-1}.
\end{align*}
The $k$th row of $\C$ can be updated by:
\begin{equation*}
\C(k,:) = \Y_3(:,k)^\top\M\left[
\M^\top\Diag(\W_3(:,k))\M+ \lambda_3\I_R\right ]^{-1},
\end{equation*}
where
\begin{equation*}
\M = [(\B_1\odot\A_1)\one_{L},\cdots,(\B_R\odot\A_R)\one_{L}].
\end{equation*}

\begin{IEEEbiography}[{\includegraphics[width=1in,height=1.25in,clip,keepaspectratio]{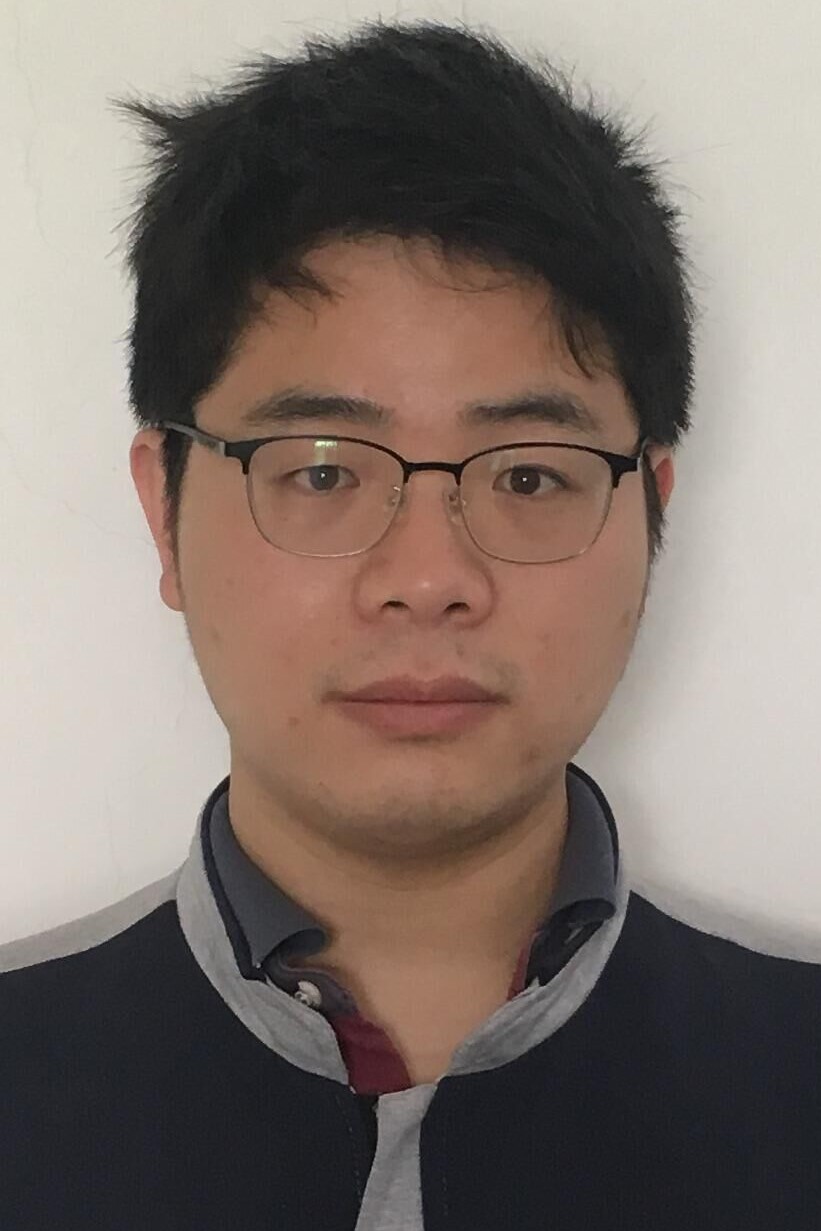}}]{Guoyong Zhang}  received the B.S. degree from the University of Electronic Science and Technology of China  (UESTC), Chengdu, China, in 2014, where he is currently working toward the Ph.D. degree with the National Key Laboratory of Science and Technology on Communications.
His currents research interests include cognitive radio and signal processing for wireless communications.
\end{IEEEbiography}

\begin{IEEEbiography}[{\includegraphics[width=1in,height=1.25in,clip,keepaspectratio]{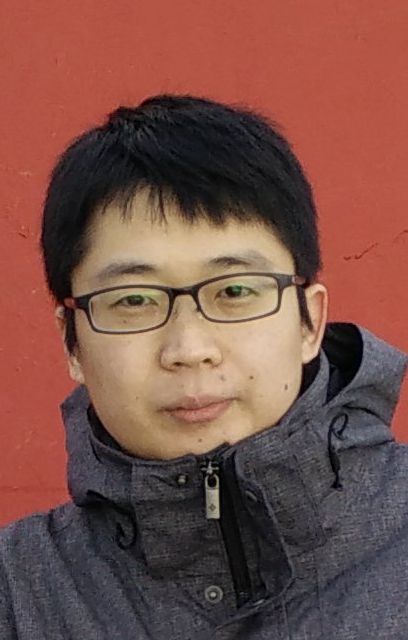}}]
                {Xiao Fu} (S'12-M'15) is an Assistant Professor in the School of Electrical Engineering and Computer Science, Oregon State University, Corvallis, Oregon, United States. He received his Ph.D. degree in Electronic Engineering from The Chinese University of Hong Kong (CUHK), Hong Kong, in 2014. He was a Postdoctoral Associate in the Department of Electrical and Computer Engineering, University of Minnesota, Minneapolis, MN, United States, from 2014 to 2017. His research interests include the broad area of signal processing and machine learning. He received a Best Student Paper Award at ICASSP 2014. Two of his co-authored papers received the Best Student Awards at IEEE CAMSAP 2015 and IEEE MLSP 2019, respectively. He also received the Outstanding Postdoctoral Scholar Award from the University of Minnesota in 2016.
He serves as the treasurer of IEEE Signal Processing Society Oregon Chapter. He also serves as a member of the EURASIP Technical Area Committee in Signal Processing for Multisensor Systems for the term of 2020-2022.
\end{IEEEbiography}

\begin{IEEEbiography}[{\includegraphics[width=1in,height=1.25in,clip,keepaspectratio]{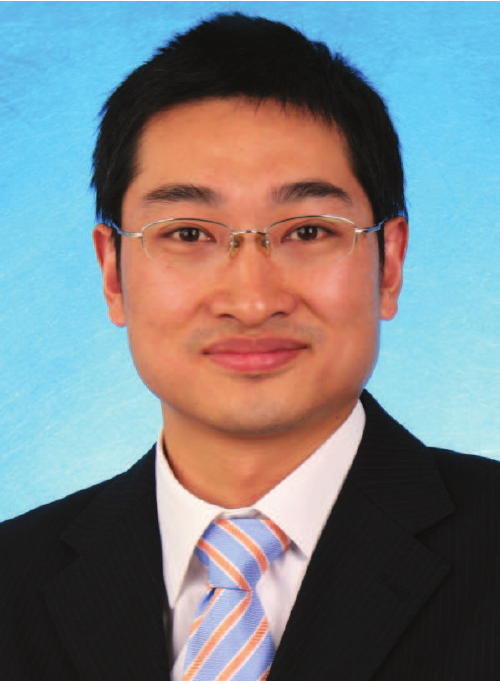}}]{Jun Wang} (S'03¨CM'09) received the B.S. degree in communication engineering and the M.S. and Ph.D. degrees in communication and information systems from the University of Electronic Science and Technology of China (UESTC), Chengdu, China, in 1997, 2000, and 2009, respectively. Since 2000, he has been with the National Key Laboratory of Science and Technology on Communications, UESTC, where he is currently a Professor. His research interests include signal processing for wireless communications, cognitive radio, machine learning for wireless communications. In these areas, he has published over 70 journal and conference papers, and involved over 10 major research programs.
\end{IEEEbiography}

\begin{IEEEbiography}[{\includegraphics[width=1in,height=1.25in,clip,keepaspectratio]{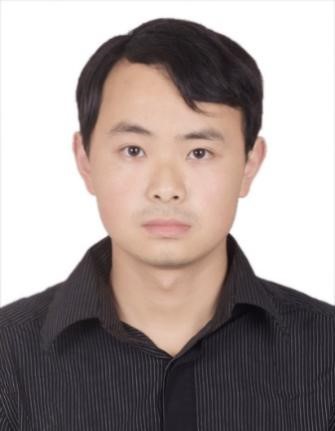}}]{Xi-Le Zhao}  received the M.S. and Ph.D. degrees from the University of Electronic Science and Technology of China (UESTC), Chengdu, China, in 2009 and 2012. He is currently a Professor with the School of Mathematical Sciences, UESTC. His research interests include image processing, computer vision, and machine learning. His website ishttps://zhaoxile.github.io/.
\end{IEEEbiography}

\begin{IEEEbiography}[{\includegraphics[width=1in,height=1.25in,clip,keepaspectratio]{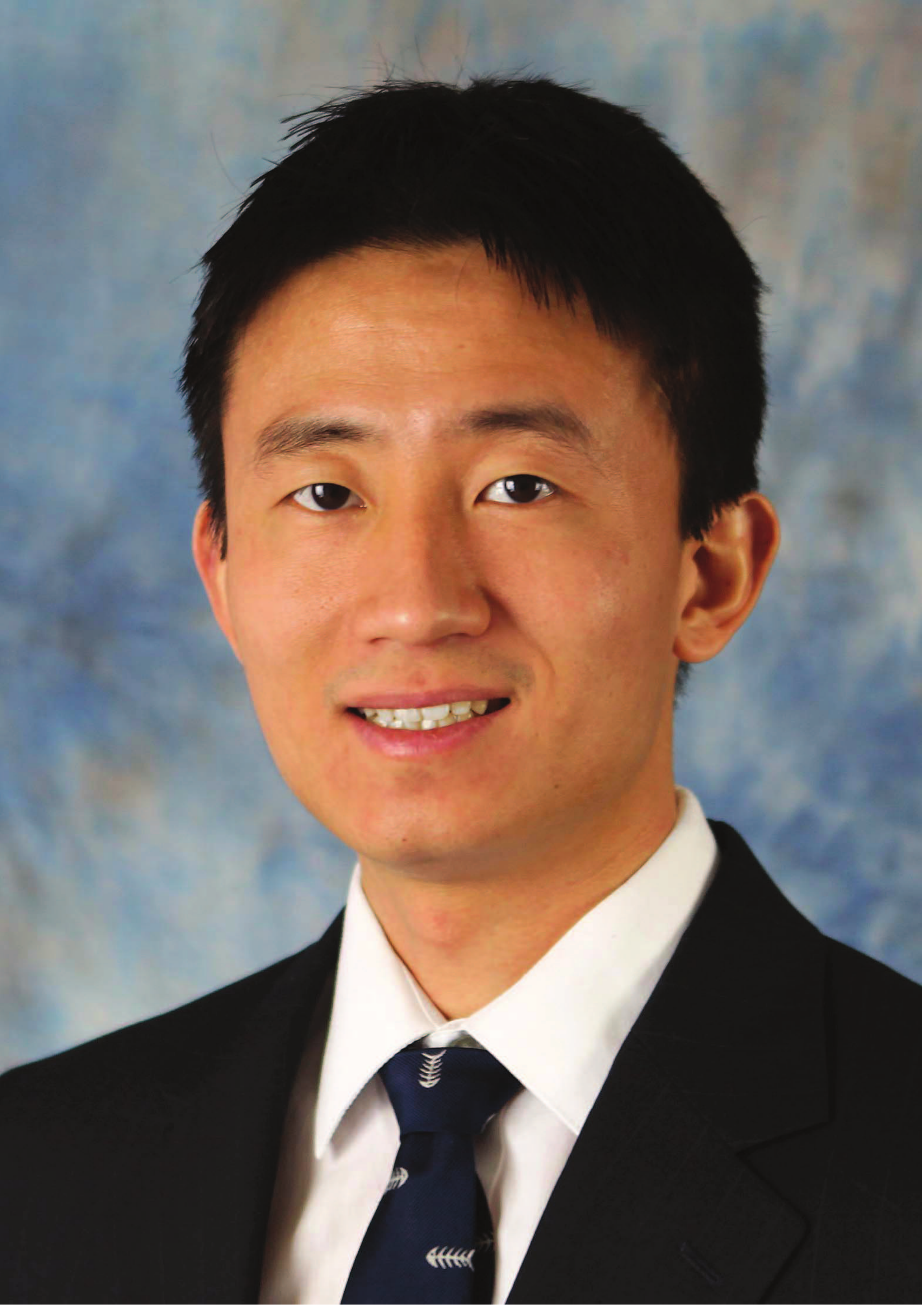}}]{Mingyi  Hong} received his Ph.D. degree from the University of Virginia, Charlottesville, in 2011. He is an assistant professor in the Department of Electrical and Computer Engineering at the University of Minnesota, Minneapolis. He serves on the IEEE Signal Processing for Communications and Networking and Machine Learning for Signal Processing Technical Committees. His research interests include optimization theory and applications in signal processing and machine learning. He is a Member of the IEEE.
Mingyi Hong was supported, in part, by the National Science Foundation (grant CIF-1910385) and Army Research Office (grant 73202-CS).
\end{IEEEbiography}

\end{document}